\documentclass[twocolumn,amsmath,amssymb,footinbib,superscriptaddress]{revtex4-2}

\usepackage{graphicx,color}
\usepackage{dcolumn}
\usepackage{bm}
\usepackage{url}
\usepackage{amsmath}
\usepackage[colorlinks=true,citecolor=blue,urlcolor=blue,linkcolor=blue]{hyperref}
\usepackage[normalem]{ulem}

\begin{document}

\title{Selective coupling of coherent optical phonons in YBa$_2$Cu$_3$O$_{7-\delta}$ with electronic transitions}

\author{Kunie Ishioka}
\affiliation{National Institute for Materials Science, Tsukuba, 305-0047 Japan}

\author{Alexej Pashkin}
\affiliation{Helmholtz-Zentrum Dresden-Rossendorf, 01328 Dresden, Germany}

\author{Christian Bernhard}
\affiliation{Department of Physics, University of Fribourg, CH-1700 Fribourg, Switzerland}

\author{Hrvoje Petek}
\affiliation{Department of Physics and Astronomy and IQ Initiative, University of Pittsburgh, Pittsburgh, PA 15260, USA}

\author{Xin Yao}
\affiliation{School of Physics and Astronomy, Shanghai Jiao Tong University, Dongchuan Road 800, 200240 Shanghai, China}

\author{Jure Demsar}
\affiliation{Institute of Physics, Johannes Gutenberg University Mainz, 55128 Mainz, Germany}

\date{\today}

\begin{abstract}

We investigate coherent lattice dynamics in optimally doped YBa$_2$Cu$_3$O$_{7-\delta}$ driven by ultrashort ($\sim$ 12 fs) near infrared (NIR) and near ultraviolet (NUV) pulses. Transient reflectivity experiments, performed at room temperature and under moderate ($<$0.1 mJ/cm$^2$) excitation fluence, reveal $A_g$-symmetry phonon modes related to the O(2,3) bending in the CuO$_2$ planes and to the apical O(4) stretching at frequencies between 10 and 15 THz, in addition to the previously reported Ba and Cu(2) vibrations at 3.5 and 4.5 THz.  
The relative coherent phonon amplitudes are in stark contrast to the relative phonon intensities in the spontaneous Raman scattering spectrum excited at the same wavelength.  
This contrast indicates mode-dependent contributions of the Raman and non-Raman mechanisms to the coherent phonon generation.  We show that the particularly intense coherent Cu(2) phonon, together with its initial phase, supports its generation predominately via a displacive mechanism, possibly involving the charge transfer within the CuO$_2$ planes. The small amplitude of the coherent out-of-phase O(2,3) bending mode at 10 THz also suggests the involvement of non-Raman generation mechanism. The generation of the other coherent phonons can in principle be explained within the framework of Raman mechanism.  When the pump light has the polarization component perpendicular to the CuO$_2$ plane, the coherent O(4) mode at 15 THz is strongly enhanced compared to the in-plane excitation, corresponding to the large polarizability component associated with the hopping between the apical and the chain oxygens, O(4) and O(1).

\end{abstract}


\maketitle

\section{INTRODUCTION}

Electron-phonon coupling in cuprate high-temperature superconductors has been examined extensively, aiming to reveal its role, or lack thereof, in the superconductivity \cite{Shen2002, He2018, Gadermaier2010, Rosenstein2021, Sugai2003, Reznik2006, Giustino2008}.  Numerous observations \cite{Shen2002} implied  a dominant role of electron-phonon coupling in the pairing \cite{Shen2002, He2018, Rosenstein2021, Nessler1998, Gweon2004, Pintschovius2005, Schrodi2021}, while strong correlations, the proximity to magnetic order \cite{Lee2006}, observations of magnetic scattering peaks \cite{Birgeneau2006} and the d-wave gap symmetry \cite{Tsuei2000, Scalapino1995} suggested the importance of short-range antiferromagnetic correlations. Moreover, the nature of electron-phonon interaction and its role on the intertwined/competing charge order instability, ubiquitous in this class of materials, has been in the research focus recently \cite{Hinton2013, Wang2021, Peng2020, Banerjee2020, Miao2018}. 
The interplay between electrons and phonons have been studied by means of conventional optical spectroscopy \cite{Sugai2003, Henn1997}, inelastic neutron and X-ray scatterings \cite{Reznik2006, Pintschovius2005, Miao2018}, photoemission \cite{Shen2002, Gweon2004}, and more recently, resonant inelastic X-ray scattering \cite{Wang2021}. In addition, the electron-phonon dynamics have been also revealed in the time domain by measuring transient reflectivity \cite{Chwalek1991, Albrecht1992, Kabanov1999, Demsar1999, Mihailovic1999, Misochko2001, Kusar2008, Novelli2017}, THz conductivity \cite{Pashkin2010, Beyer2011}, time-resolved photoemission \cite{Nessler1998, Smallwood2012, Cilento2018}, diffraction \cite{Gedik2007, Carbone2008}, Raman scattering \cite{Mertelj1997, Saichu2009} as well as by means of THz excitation of infrared-active lattice vibrations \cite{Hu2014, Mankowsky2015, Liu2020}. 

\begin{figure}
\includegraphics[width=0.475\textwidth]{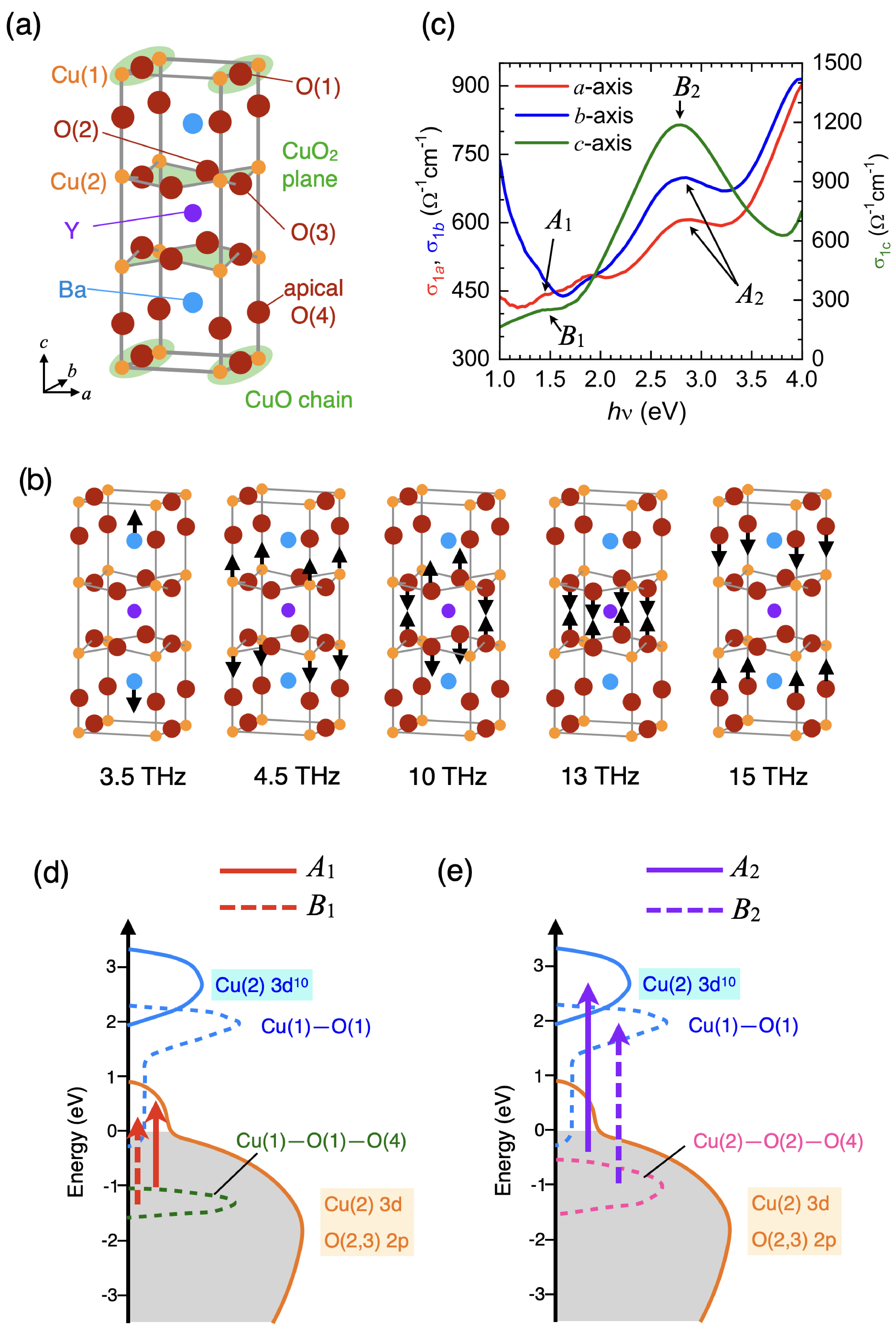}
\caption{\label{Band}  
 (a) Crystalline structure of orthorhombic YBa$_2$Cu$_3$O$_{7-\delta}$.  
 (b) Schematic illustration of the ionic motions of the $A_g$ phonon modes in orthorhombic YBa$_2$Cu$_3$O$_{7-\delta}$ \cite{Liu1988, Bates1989}.
 (c) Optical conductivity spectra for light polarized along the $a, b$ and c axes of a detwinned crystal. (d, e) Schematic illustrations of the density-of-states (DOS) of YBa$_2$Cu$_3$O$_{7}$ near the Fermi level and possible transitions with photon energy of 1.5 eV (d) and 3 eV (e) \cite{Romberg1990}. The solid (dashed) arrows and curves represent possible transitions for light polarized in the ab-plane (along the c axis) and the DOS associated with them.  Occupied states are shaded with grey.
}
\end{figure}

YBa$_2$Cu$_3$O$_{7-\delta}$ is among the high-temperature superconductors whose lattice vibrations and their coupling to electronic degrees of freedom have been investigated most extensively.  
For optimally doped ($\delta<$0.1) orthorhombic YBa$_2$Cu$_3$O$_{7-\delta}$, whose crystalline structure is shown in Fig.~\ref{Band}a, Raman active vibrations are those of Ba, Cu(2), O(2), O(3) and O(4) ions, each of which gives rise to three Raman active modes of $A_g+B_{2g}+B_{3g}$ symmetries  \cite{Kulakovskii1988}.  When excited and detected with the light polarized within the crystallographic ab plane, the Raman spectra comprise five $A_g$-symmetry modes whose ionic displacements are all along the crystallographic c axis, as shown in Fig.~\ref{Band}b \cite{Burns1988, Macfarlane1988, Thomsen1988, Liu1988, Heyen1990, McCarty1990, Hong2010}.  The corresponding irreducible representation of the Raman tensor is given by:
\begin{equation}\label{Rtensor}
\mathfrak{R}(Z)= \begin{pmatrix}
a_{xx} & 0 & 0\\
0 & a_{yy} & 0 \\
0 & 0 & a_{zz}
\end{pmatrix},
\end{equation}
for the ionic motion along the $z$ direction.  The intensity of each phonon mode is affected differently upon varying temperature across the superconducting transition temperature $T_c$ \cite{Macfarlane1987, Friedl1991, Altendorf1993, Henn1997, Limonov1998} as well as upon varying excitation photon energy \cite{Heyen1990, Heyen1992}. Moreover, their frequencies also depend on the oxygen content (i.e., in-plane hole density) \cite{Kourouklis1987, Macfarlane1988, Thomsen1988, Opel1999, Hong2010}.  These  dependencies imply selective coupling of electronic states to distinct ionic motions in YBa$_2$Cu$_3$O$_{7-\delta}$.

Raman-active phonon modes can be studied also as periodic temporal modulations in transient transmissivity or reflectivity following photoexcitation with a femtosecond laser pulse.   This scheme is powerful in examining the coupling between photoexcited particles and the lattice degrees of freedom, especially when combined with a broadband probe pulse and theoretical calculations \cite{DalConte2012, Mansart2013, Novelli2017, Baldini2017, Baldini2020}.  Two representative generation mechanisms have been put forward:  impulsive stimulated Raman scattering (ISRS) and the displacive excitation of coherent phonons (DECP).   In the ISRS the pump electric field $E$ acts directly upon the ions in the crystal through the Raman tensor $\mathfrak{R}\equiv(\partial\chi/\partial Q)_{jkl}$ \cite{Dhar1994, Merlin1997, Dekorsy, Mann2015}, as elaborated in Appendix~\ref{ISRS}.  By contrast, the DECP driving force is generally given by the dependence of the equilibrium coordinate $Q_E$ on excited carrier density $N$ (or electronic temperature $T_\text{e}$) \cite{Zeiger1992}, as described in Appendix~\ref{DECP}.

Earlier time-resolved experiments on YBa$_2$Cu$_3$O$_{7-\delta}$ with a single-color or broadband probe \cite{Chwalek1991, Albrecht1992, Misochko1999, Misochko2000, Misochko2001, Misochko2002, Takahashi2009, Hinton2013, Dakovski2015, Mankowsky2015, Novelli2017, Ramos2019} reported the two lowest-frequency $A_g$ modes associated with the Ba and Cu(2) displacements.  These two modes exhibited qualitatively different dependencies on temperature across $T_c$ \cite{Albrecht1992, Misochko2000, Misochko2002, Novelli2017}. 
The Ba amplitude was enhanced considerably below $T_c$, whereas the Cu(2) amplitude was independent of temperature.  While these temperature-dependencies were qualitatively similar to those of the Raman intensities of the two modes \cite{Friedl1991},  early theoretical studies \cite{Mazin1994, Nettel1994} proposed that the coherent Ba phonon in the superconducting state is generated via a mechanism similar to DECP.  
The originally proposed DECP was associated with the Peierls instability of group V semimetals \cite{Zeiger1992}. In YBa$_2$Cu$_3$O$_{7-\delta}$, however, it was argued that superconductivity is accompanied by a static displacement in $Q_E$ of the Ba ion, and that photo-induced suppression of superconductivity leads to a sudden shift in $Q_E$ and thereby launches coherent Ba mode \cite{Mazin1994, Nettel1994}.  
On the other hand, the experimentally observed initial phases of the coherent phonons \cite{Misochko2000, Misochko2001, Misochko2002} as well as the result of a double-pump/probe experiment  \cite{Lobad2001} argued against the DECP mechanism.

In the normal state, by contrast, there has been very little discussion on the generation mechanism for the Ba and Cu(2) modes beyond the implicit assumption of the standard ISRS mechanism.
For the higher-frequency phonons 
there have only been a few time-resolved 
studies 
\cite{Misochko2001, Ramos2019}, and little has been known about their coupling with non-equilibrium carriers, either in the normal state or in the superconducting state.
 
In optimally doped YBa$_2$Cu$_3$O$_{7-\delta}$, the electronic excitations have a relatively localized character and depend critically on the photon energy as well as on the polarization of the light. 
The optical conductivity spectra, shown in Fig.~\ref{Band}c, exhibit a weak peak at $\sim$1.5~eV ($A_1$) and a stronger peak at $\sim$2.75 eV ($A_2$) for the light polarized parallel to the ab (or CuO$_2$) plane. The $A_1$ and $A_2$ peaks have been linked to the localized $d-d$ excitations within the broad, partially filled O 2$p$ -- Cu 3$d$ band and to the optical transition to the unoccupied Cu 3$d^{10}$ band, respectively, as shown by solid arrows in the schematic density-of-states (DOS) in Fig.~\ref{Band}d and e \cite{Romberg1990, Kircher1992, Heyen1992}. While both $A_1$ and $A_2$ transitions involve the in-plane charge transfer between O(2,3) and Cu(2) ions, the transition $A_2$ has a significantly larger dipole moment.
For the light polarized parallel to the c axis, i.e., perpendicular to the CuO$_2$ plane, the optical conductivity spectra in Fig.~\ref{Band}c exhibit the peaks $B_1$ and $B_2$ that correspond to the excitation from occupied Cu(1)-O(1)-O(4) band into the unoccupied Cu(1)-O(1) chain band and the plane-to-chain charge transfer \cite{Kircher1991, Kircher1992}, respectively, as shown by the dashed arrows in Fig.~\ref{Band}d and e. These excitation schemes imply that the resonant electronic transitions are localized within the crystalline unit cell of YBa$_2$Cu$_3$O$_{7-\delta}$, and suggest the possibility of selective electron-phonon coupling contributing to the coherent phonon generation.

In this study we extend the range of accessible coherent phonons to all the $A_g$ phonons depicted in Fig.~\ref{Band}(b) 
by using ultrashort ($\lesssim$ 12~fs) optical pulses and investigate their coupling with different optical transitions by tuning the photon energy and optical polarization. 
We perform transient reflectivity measurements on optimally doped YBa$_2$Cu$_3$O$_{7-\delta}$ single crystal excited with near infrared (NIR) and near ultraviolet (NUV) light at energies of 1.5 and 3.1 eV, which roughly match the resonant $A_1/B_1$ and $A_2/B_2$ transitions
. Comparison of the relative coherent phonon amplitudes with the spontaneous Raman scattering intensities reveal a pronounced difference and indicates 
that Cu(2) mode is generated predominantly by a DECP mechanism. The phonon amplitudes in the NIR experiment depend also on the polarization of the pump 
light, demonstrating particularly strong coupling of the optical transition to the Cu-O chain band with the O(4) mode at 15 THz. 
 
\section{EXPERIMENTAL}

A single crystal of YBa$_2$Cu$_3$O$_{7-\delta}$ was grown by the top-seeded solution growth \cite{Yao1997}. Prior to optical measurements the sample was cleaved along the ab-plane and annealed in flowing oxygen at 500$^\circ$C for 72 hours to obtain optimal doping ($\delta\simeq$0.05) \cite{Lindemer1989}. Investigation under the polarization microscope revealed twinning domains with typical length of 20-50 $\mu$m and width of 2-3 $\mu$m within the ab-plane \cite{Lin2004, Mei2005}.

Single color pump-probe reflectivity measurements were performed under ambient conditions at room temperature. For NIR pump-probe study we used the fundamental output of a Ti:sapphire oscillator (Griffin, KM Labs) with an 80 MHz repetition rate, a 12-fs duration and a broadband spectrum 
at 815$\pm$65 nm (1.53$\pm$0.12 eV photon energy). For NUV experiments a second harmonic of an 80-MHz repetition-rate 
Ti:sapphire oscillator was used with 10~fs pulses at 394$\pm$17 nm (3.16$\pm$0.13 eV). 
In both experiments the 
probe light is polarized along either a or b axis of the twinned crystal.
In the NIR back-reflection experiments the pump and probe lights are polarized perpendicular to each other, whereas in the NUV they are polarized parallel.
The optical penetration depth of the NIR and NUV lights in YBa$_2$Cu$_3$O$_{7-\delta}$ is $1/\alpha \approx~$90 and 70 nm. The pump and probe laser spot size on the sample was $\gtrsim$60 and $\gtrsim$30 $\mu$m in diameter in the normal-incidence NIR and NUV experiments. 
The pump-induced change in reflectivity $\Delta R$ is measured  by detecting the probe light before and after reflection at the sample surface with a pair of matched photodetectors. The differentially detected signal from the photodetector pair is pre-amplified and averaged on a digital oscilloscope, typically over 10,000 scans, while the time delay $t$ between the pump and probe is scanned at 20 Hz.
	
To compare the relative amplitudes of coherent phonons with their Raman intensities, Raman scattering spectra were measured in the back-scattering geometry using a Renishaw inVia Raman microscope with excitation at 785 nm.  The laser spot was an ellipsoid with $\sim5\times30~\mu$m$^2$ size.  

The linear optical response function of a detwinned YBa$_2$Cu$_3$O$_{6.95}$ crystal, shown in Fig.~\ref{Band}b, was measured in the NIR 
to NUV 
range with a spectroscopic ellipsometer in rotating compensator configuration. The measurements were performed in three different geometries for which the plane of incidence was directed along either the a, b, or c axis of the detwinned crystal. The obtained spectra were corrected for anisotropy effects using the standard Woollam software to obtain the a-, b- and c-axis components of the optical conductivity.

\section{RESULTS AND DISCUSSION}

\subsection{Normal-incidence NIR and NUV pulse excitations}\label{NormalIncidence}

We first examine the transient reflectivity dynamics using NIR and NUV pulses in the near normal-incidence geometry, where the pump and probe beams are polarized parallel to the ab plane. In this configuration the dominant electronic transition occurs within the partially filled Cu(2) 3$d$ and O(2,3) 2$p$ bands for the NIR excitation and from the Cu(2) 3$d$ and O(2,3) 2$p$ band into the higher lying Cu(2) 3$d^{10}$ for the NUV excitation, as shown by the solid arrows in Fig.~\ref{Band}d,e.  
The ISRS driving force for the coherent $A_g$ phonons, which is given by Eq.~(\ref{A5}) in Appendix~\ref{ISRS}, would reduce to:
\begin{equation}\label{force3}
F^\textrm{ISRS}_z(t)=a_{xx}|E_x|^2+a_{yy}|E_y|^2,
\end{equation}
where $E_x$ and $E_y$ denote the components of the pump electric field $E$ along the crystallographic a and b axes.

\begin{figure}
\includegraphics[width=0.475\textwidth]{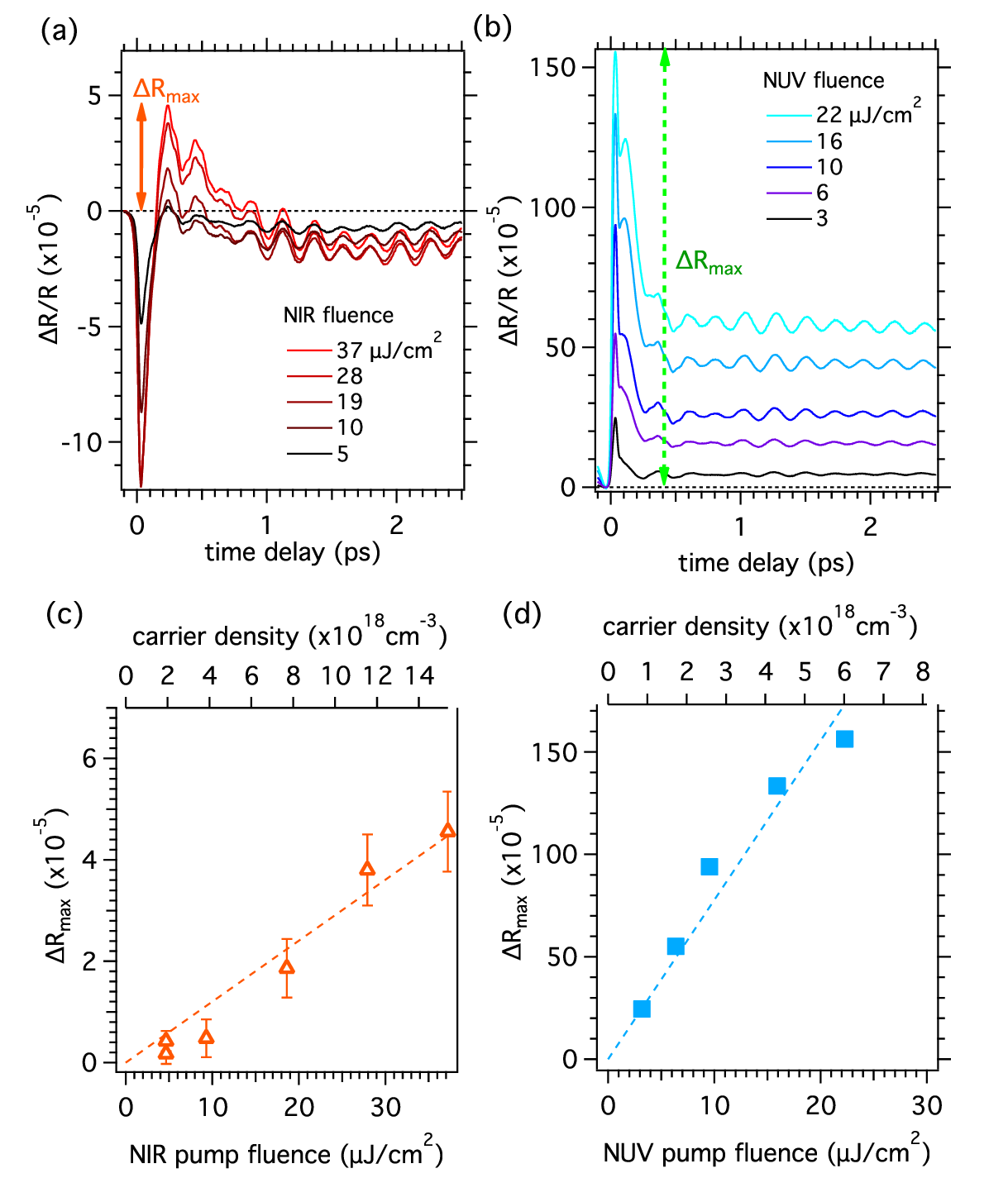}
\caption{\label{ElectronicPower} (a,b) Transient reflectivity of YBa$_2$Cu$_3$O$_{7-\delta}$ obtained at different incident pump fluences in the normal-incidence geometry using NIR (a) and NUV (b) pulses. The peak height $\Delta R_\text{max}$ in reflectivity transient immediately after photoexcitation are indicated by arrows in both panels. (c,d) Pump fluence-dependence of the peak height $\Delta R_\text{max}$ in the NIR (c) and NUV (d) measurements. The error bars 
reflect the amplitude of the most prominent oscillatory component, which is due to the coherently driven Ba mode.  Lines are to guide the eye.}
\end{figure}

Figure~\ref{ElectronicPower}a,b presents the transient reflectivity traces at different incident pump fluences. The NIR signals in Fig.~\ref{ElectronicPower}a show an instantaneous drop at $t\simeq 0$, followed by a rapid rise to reach a maximum at $t\simeq 250$ fs and then by a slower decay to a negative baseline. 
The height of the reflectivity maximum, $\Delta R_\text{max}$, increases roughly linearly up to $\gtrsim$30 $\mu$J/cm$^2$, as shown in Fig.~\ref{ElectronicPower}c. The incoherent response can be attributed to the creation of hot electron gas and its subsequent cooling on a sub-picosecond time scale via thermalization with the lattice \cite{Kabanov1999, Perfetti2007, Gadermaier2010, DalConte2015}.  
In the NUV measurements presented in Fig.~\ref{ElectronicPower}b, the transient reflectivity  reaches its maximum at an earlier time delay ($t\sim$ 60 fs) and decays faster than in the NIR measurements. $\Delta R_\text{max}$ for the NUV experiments also exhibits a roughly linear fluence-dependence, as shown in Fig.~\ref{ElectronicPower}d. Here, the response can be interpreted as the excitation of electrons into the higher lying Cu(2) 3$d^{10}$ band, followed by their thermalization and cooling.  

\begin{figure}
\includegraphics[width=0.475\textwidth]{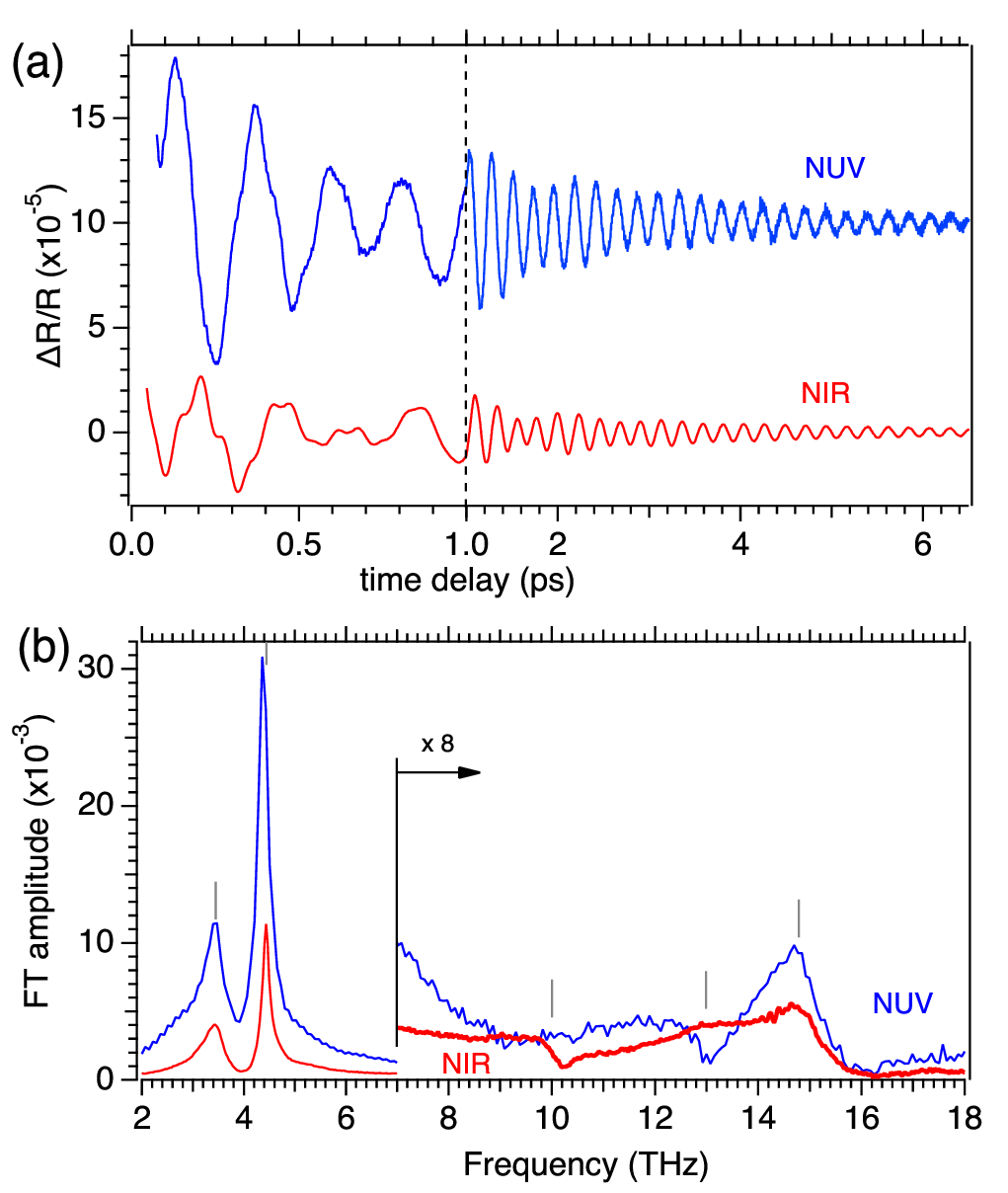}
\caption{\label{TDFT800_400} 
Oscillatory parts of transient reflectivity (a) and their fast Fourier-transform (FFT) spectra (b) of YBa$_2$Cu$_3$O$_{7-\delta}$ obtained in the normal-incidence NIR and NUV pump-probe configurations. Pump fluences are 19 and 22 $\mu$J/cm$^2$ for the NIR and NUV measurements, corresponding to the photoexcited carrier density of $\approx$$8\times10^{18}$ and $6\times10^{18}$ cm$^{-3}$, respectively. Traces in (a) are offset for clarity. Gray vertical lines in (b) represent approximate frequencies of the $A_g$-symmetry phonons reported in the previous Raman studies \cite{McCarty1990, Schutzmann1995}.
}
\end{figure}

On top of the electronic response, an oscillatory modulation of the reflectivity by coherent phonons is clearly resolved in both NIR and NUV experiments. For the NIR experiment, the oscillatory part, shown in Fig.~\ref{TDFT800_400}a, can be relatively uniquely separated from the non-oscillatory electronic response by approximating the latter with a multi-exponential function.  For the NUV experiment, the separation at early delay times ($t\lesssim0.2$) ps is more difficult due to the large positive spike and the following hump in the electronic response.
Fast Fourier transform (FFT) spectra of the oscillatory reflectivity are shown in Fig. ~\ref{TDFT800_400}b.

\begin{figure}
\includegraphics[width=0.475\textwidth]{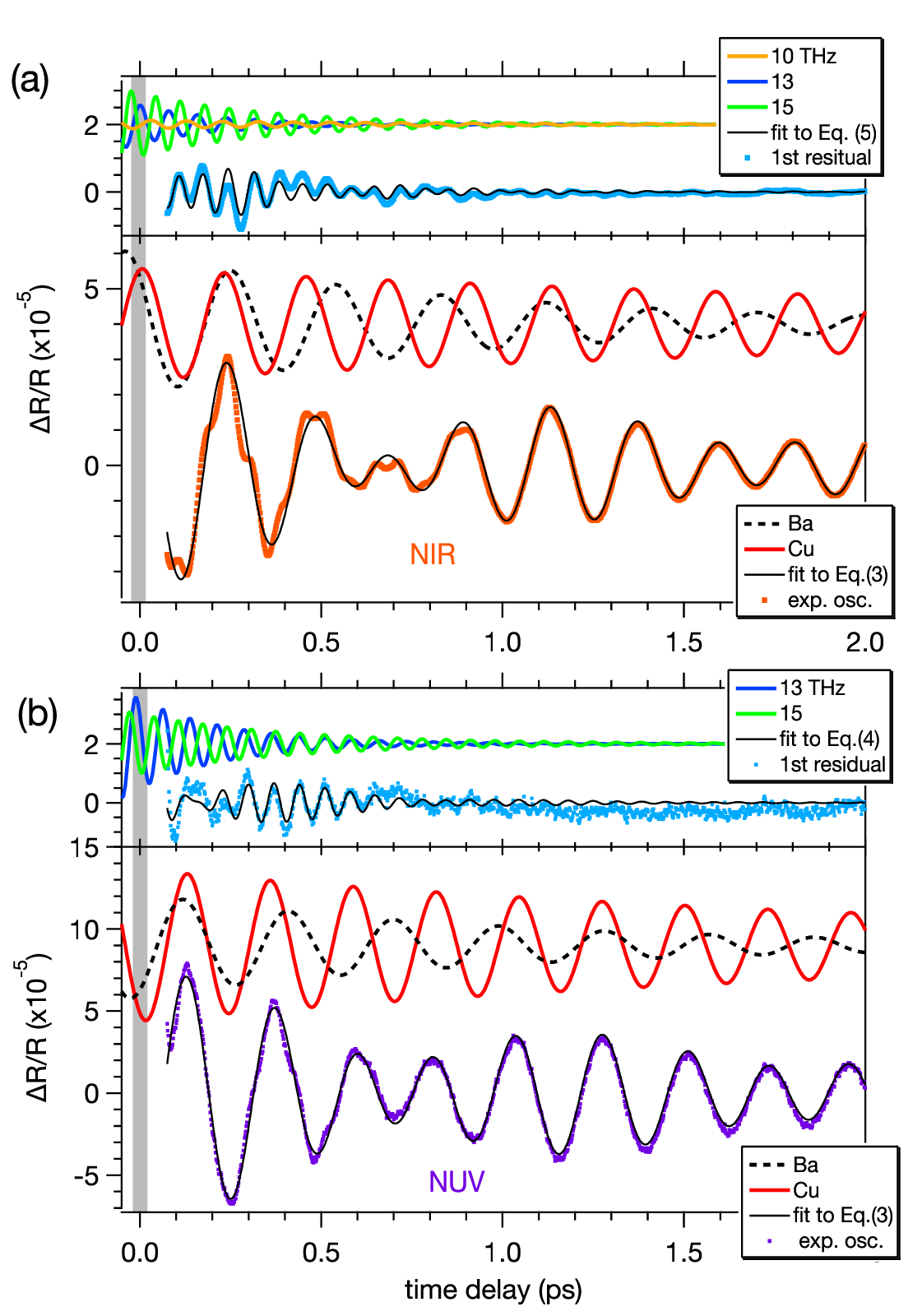}
\caption{\label{fit} 
Oscillatory transient reflectivity (dots in bottom panels) obtained with normal incidence NIR (a) and NUV (b) configurations.  Black solid curve, black dashed curve and red solid curve in bottom panels are fit to Eq.~(\ref{ddh}) and its Ba- and Cu-mode contributions, with the latter two being offset for clarity.  Dots in the upper panels represent the residuals of the first fitting shown in the bottom panels.  Black curve is the fit to Eq.~(\ref{mdh}) or (\ref{mdh3}), with colored curves representing the contributions from the 10-, 13- and 15-THz modes offset for clarity. Gray shades represent the upper limit of the uncertainty in determining $t=0$.}
\end{figure}

For both experiments, the oscillatory part for $t\geq1$ ps can be well reproduced by two damped harmonic oscillators:
\begin{eqnarray} 
\label{ddh}
f(t)&=&A_\text{Ba} \exp(-\Gamma_\text{Ba} t)\sin(2\pi\nu_\text{Ba} t+\phi_\text{Ba})\nonumber\\
&+&A_\text{Cu} \exp(-\Gamma_\text{Cu} t)\sin(2\pi\nu_\text{Cu} t+\phi_\text{Cu}),
\end{eqnarray}
%
as shown in the lower panels of Fig.~\ref{fit}a,b.  Here the subscripts denote the coherent Ba and Cu(2) modes, which were reported at $\sim$3.5 and 4.5 THz in the previous time-resolved \cite{Chwalek1991, Albrecht1992, Misochko1999, Misochko2000, Misochko2001, Misochko2002, Takahashi2009, Hinton2013, Dakovski2015, Mankowsky2015, Novelli2017, Ramos2019} and spontaneous Raman scattering \cite{Macfarlane1988, McCarty1990, Henn1997, Hong2010, Krol1987, Liu1988, Heyen1990, Altendorf1993, Kourouklis1987, Burns1988, Thomsen1988, Bates1989} studies.  Their ionic displacements are illustrated in Fig.~\ref{TDFT800_400}c.  The frequencies $\nu_j$ obtained from the fitting of the NIR and NUV data are listed in the first and second columns of Table~\ref{CPAg}.  The uncertainty in determining the frequency is $\lesssim0.1$ THz, including the fluence-dependent frequency shifts shown in Appendix~\ref{Heating}.
These two modes also dominate the low-frequency regime of the 
FFT spectra shown in Fig.~\ref{TDFT800_400}b.

\begin{table*}
\caption{\label{CPAg} Phonon frequencies in THz (cm$^{-1}$) as observed in the transient reflectivity measurements in normal- and oblique-incidence configurations, compared to 
spontaneous Raman 
peak frequencies obtained in this work and in literatures.
The uncertainty in determining the frequency is $<0.1$ THz for the Ba and Cu modes  and $<0.5$ THz for the high-frequency modes, both including the fluence-dependent frequency shifts.}
\begin{ruledtabular}
\begin{tabular}{rrrrcl}
NIR normal&NUV normal&NIR oblique&Raman NIR&Literature&Assignment and symmetry\\
THz (cm$^{-1})$&THz (cm$^{-1}$)&THz (cm$^{-1}$)&(cm$^{-1}$)&(cm$^{-1}$)&\\
\hline
3.44 (115)&3.46 (115)&3.49 (116)&113&114-116\footnote{Ref.~\cite{McCarty1990}}&Ba symmetric stretch $\parallel$c, $A_g$
\\
4.43 (148)&4.37 (146)&4.46 (149)&151&149$^{a}$&Cu(2) symmetric stretch $\parallel$c, $A_g$\\
10.1 (338)&&10.1 (338)&337&335-336$^a$&O(2,3) out-of-phase bend $\parallel$c, $A_g$\footnote{The out-of-phase O(2,3) bending mode at 10 THz has $A_g$ symmetry in the orthorhombic phase ($0<\delta<0.5$) and $B_{1g}$ symmetry in the tetragonal phase ($0.5<\delta<1$).} \\
12.7 (423)&13.1 (437)&12.8 (429)&&435$^a$&O(2,3) in-phase bend $\parallel$c, $A_g$\\
14.8 (495)&14.7 (489)&14.9 (497)&493&493-498$^a$&O(4) symmetric stretch $\parallel$c, $A_g$\\
&&&577&575\footnote{Ref.~\cite{Liu1988}}&O(1) stretch $\parallel$b, $B_{2u}$\\
\end{tabular}
\end{ruledtabular}
\end{table*}

For the earlier time delays ($t<1$ ps), the extrapolation of the fitting to Eq.~(\ref{ddh}) reproduces the oscillatory reflectivity approximately but not perfectly.  The residuals of the first fit to Eq.~(\ref{ddh}), shown in the upper panels of Fig.~\ref{fit}a,b, exhibit oscillations at apparently higher frequencies than the Ba and Cu(2) modes.  These oscillations are responsible for the weak peaks and dips appearing between 8 and 18 THz in the FFT spectra in Fig.~\ref{TDFT800_400}b.  In this frequency regime we expect to observe three $A_g$-symmetry modes that were reported in the previous Raman studies but have not been reported in the time-resolved studies: the out-of-phase and in-phase O(2,3) bending modes at $\sim$340 and 440 cm$^{-1}$, and the O(4) stretching mode at $\sim$500 cm$^{-1}$.  Indeed, in the NIR experiment one can recognize a peak-dip structure in the FFT spectrum at $\sim$10 THz in addition to peaks at $\sim13$ and 15 THz (Fig.~\ref{TDFT800_400}b), in good agreement with the Raman frequencies in the literature \cite{McCarty1990, Schutzmann1995}.  For the NUV experiment, by contrast, the 10-THz peak cannot be clearly resolved.  

The lineshapes of the high-frequency modes, i.e., whether they appear as a peak, a dip or a bipolar structure, are extremely sensitive to the choice of the temporal window employed for the FFT, as demonstrated 
in Appendix~\ref{TimeWindow}.  This is due to the interference among the strongly damped high-frequency phonon modes. We therefore determine the parameters of the high-frequency modes by further fitting the first residual in the time domain.

 For both the NIR and NUV experiments, the first residual can be reproduced roughly by a beating pattern between two damped harmonic oscillators with frequencies at around 13 and 15 THz:  
\begin{eqnarray}\label{mdh}
g(t)&=&A_{13} \exp(-\Gamma_{13} t)\sin(2\pi\nu_{13} t+\phi_{13})\nonumber\\
&+&A_{15} \exp(-\Gamma_{15} t)\sin(2\pi\nu_{15} t+\phi_{15})
\end{eqnarray}
with the subscripts denoting the 13- and 15-THz modes.  We assume the frequencies to be independent of $t$, though the recent time-resolved Raman scattering study, recorded at a fluence that is over two-orders-of-magnitude higher than in our case, reported a transient blueshift of the 15-THz mode over the first picosecond \cite{Pellatz2021}.  For the NIR experiment, fitting the first residual to Eq.~(\ref{mdh}) leaves a second residual, which is shown 
in Appendix~\ref{2vs3}.  This second residual features a small but distinct oscillatory component at around 10 THz, seen also in the FFT spectrum (red trace in Fig.~\ref{fit2}b).  We therefore add a third damped harmonic component at around 10 THz to Eq.~(\ref{mdh}) to fit the NIR data:
\begin{eqnarray}\label{mdh3}
g(t)&=&A_{10} \exp(-\Gamma_{10} t)\sin(2\pi\nu_{10} t+\phi_{10})\nonumber\\
&+&A_{13} \exp(-\Gamma_{13} t)\sin(2\pi\nu_{13} t+\phi_{13})\nonumber\\
&+&A_{15} \exp(-\Gamma_{15} t)\sin(2\pi\nu_{15} t+\phi_{15}),
\end{eqnarray}
which provides a better fitting result. In the NUV experiment, by contrast, adding the 10-THz term to the fit function does not lead to a reasonable convergence. The frequencies obtained from the multi-mode fitting of the NIR data are listed in the first and second columns of Table~\ref{CPAg}.  The uncertainty in the extracted frequencies is $<0.5$ THz for the high-frequency modes.

 We note that in the present study we do not resolve the 14-THz mode, which was observed under high excitation density in a previous transient reflectivity study \cite{Ramos2019} and was attributed to a third harmonic of the Cu(2) mode at 4.5 THz.

\subsection{Dependence on pump fluence}\label{subsc_fluence}

\begin{figure}
\includegraphics[width=0.475\textwidth]{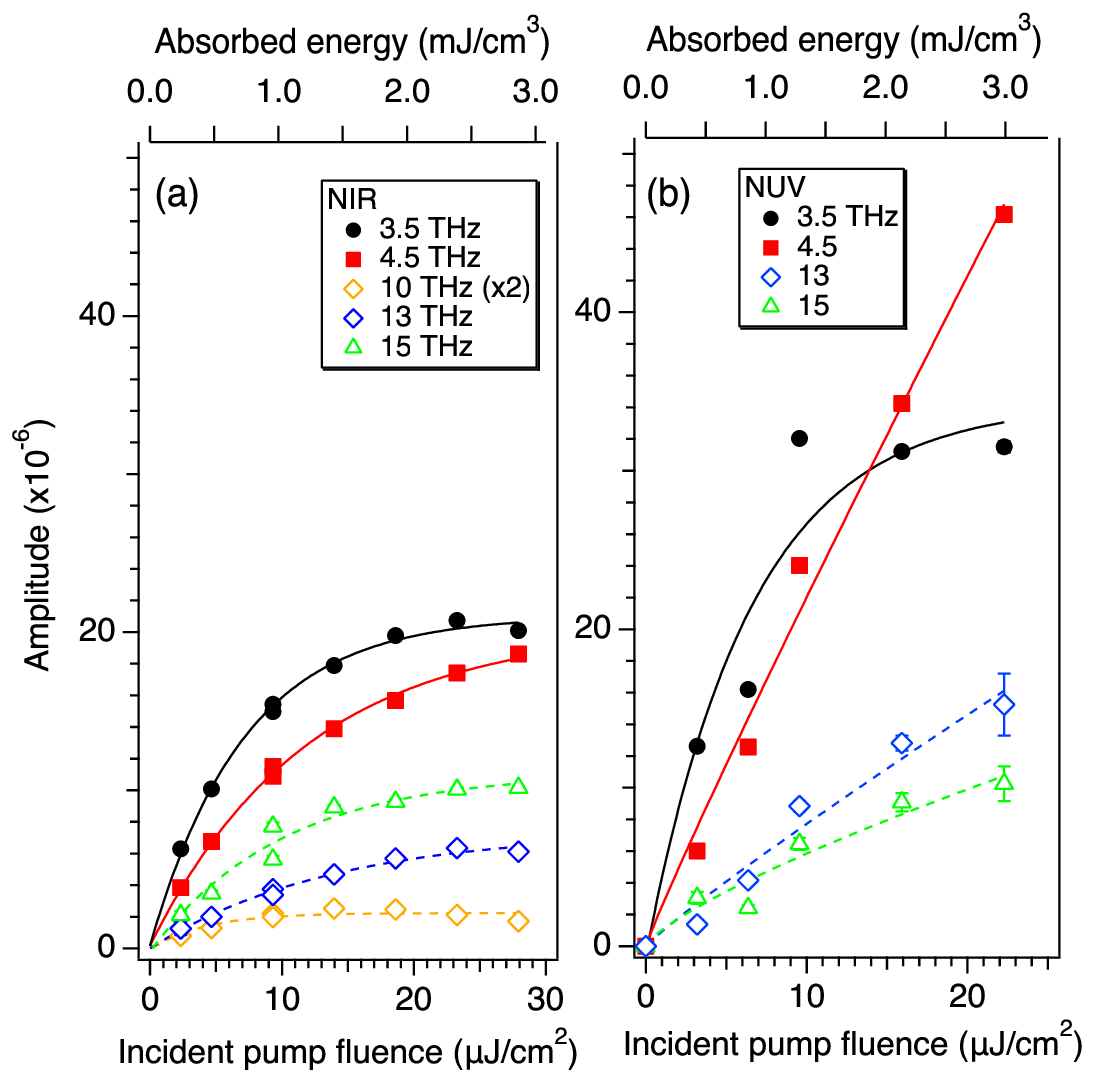}
\caption{\label{Power}  Pump fluence-dependence of the initial amplitudes of the coherent phonons for the normal-incidence NIR (a) and NUV (b) configurations obtained by fitting the time-domain oscillations as described in the main text.  For the NUV pumping the 10~THz mode was too weak to be fitted confidently.  Lines are to guide the eye. 
}
\end{figure}

The amplitudes $A_j$ of the coherent phonon modes are also obtained by the fitting procedure described above.
Figure~\ref{Power} presents the amplitudes as a function of incident pump fluence $\Phi$ (bottom axis) and absorbed photon energy density $(1-R)\Phi\alpha$ (top axis).  
Here the dephasing rates $\Gamma_j$ and frequencies $\nu_j$ of the 10- and 13-THz modes were fixed to the values obtained at the highest pump fluence (Fig.~\ref{fit}) in order to obtain a consistent result throughout the examined fluence range.  This approach is justified by the rather weak fluence-dependence of the frequencies and dephasing rates of the Ba and Cu modes (
Appendix~\ref{Heating}), which are obtained from restriction-free fitting.

In the NIR experiment the amplitudes tend to increase sub-linearly with increasing pump fluence.  The observation is in contrast to the fluence-dependence of the electronic response $\Delta R_\text{max}$ shown in Fig.~\ref{ElectronicPower}c, which indicates that the photoexcited carrier density depends linearly on the fluence.  
The overall sub-linear dependencies may partly be a result of continuous laser heating in the present high-repetition laser experiment, as discussed in Appendix~\ref{Heating}. However, the 3.5-THz mode exhibits a particularly strong sub-linearity compared with the other four modes, with an apparent saturation at $\Phi\simeq20 \mu$J/cm$^2$ or $(1-R)\Phi\alpha\simeq2$ mJ/cm$^3$.
Because the driving force for the ISRS generation (Eq.~(\ref{force3})) is proportional to the pump light fluence, the mode-dependent fluence-dependencies may be an indication of a non-Raman generation mechanism.

In the NUV experiment the phonon amplitudes increase almost linearly, except for the Ba mode that shows a sub-linear dependence and apparent saturation.  The linear fluence-dependence for the Cu(2), 13- and 15-THz modes suggests that continuous laser heating is not suppressing the  excitation of these modes, although their fluence-dependent frequencies and the dephasing rates (
Appendix~\ref{2vs3}) indicate a moderate temperature rise.
If we compare the NIR and NUV results at the same absorbed energy in the ``linear" regime, e.g., at $(1-R)\Phi\alpha=0.5$ mJ/cm$^{3}$, the phonon amplitudes for the 3.5-, 4.5-, 13- and 15-THz modes are comparable between the two measurements.  
This observation is qualitatively consistent with the previous resonant Raman study at low temperatures \cite{Heyen1990}, where all the $A_g$ phonon modes showed moderate photon energy-dependence between 1.8 to 2.7~eV.
At a higher absorbed energy e.g., at 3.0 mJ/cm$^{3}$, however, the amplitudes for the 3.5-, 4.5- and 13-THz modes in the NUV experiment are considerably larger than those in the NIR experiment.

We note that the fluence-dependences observed in the present NIR experiments (Fig.~\ref{Power}a) are qualitatively similar to those reported for the superconducting state YBa$_2$Cu$_3$O$_7$ in the earlier time-resolved study \cite{Novelli2017}. There, the 3.5-THz mode amplitude showed a saturation at $\sim20~\mu$J/cm$^2$, whereas the 4.5-THz mode saturates at about 5 times higher excitation density.  Such a saturation could in general be associated with optically induced suppression of superconductivity \cite{Pashkin2010, Beyer2011}.   In Ref.~\onlinecite{Novelli2017} the different fluence-dependences, together with the contrasted temperature-dependences, lead to an argument that the Ba and Cu(2) modes couple with the delocalized superconducting states and with the more localized electronic excitation, respectively.   
In the present normal state study, however, it is not likely that the saturation of the Ba mode is linked to quenching of some underlying order.

\subsection{Comparison with spontaneous Raman scattering and generation mechanism}\label{CompRaman}

To further examine the electron-phonon coupling behind the coherent phonon generation we measure a spontaneous Raman spectrum of the same  sample at the similar NIR excitation photon energy.  If coherent phonons are generated via ISRS, the phonon-induced reflectivity modulation $\Delta R$ would be proportional to the product of the polarizability component defined by the pump polarization $a_{kl}$ and that defined by the probe polarization $a_{pq}$ (Appendix~\ref{detect}). Spontaneous Raman scattering intensity $I_s$ is proportional to the polarizability squared $a_{kl}^2$ (Appendix~\ref{SpRaman}).  With appropriate selection of light polarizations we could therefore examine the generation mechanism based on whether or not the \emph{relative} amplitudes of the coherent $A_g$ modes agrees quantitatively with their relative Raman intensities.

\begin{figure}
\includegraphics[width=0.475\textwidth]{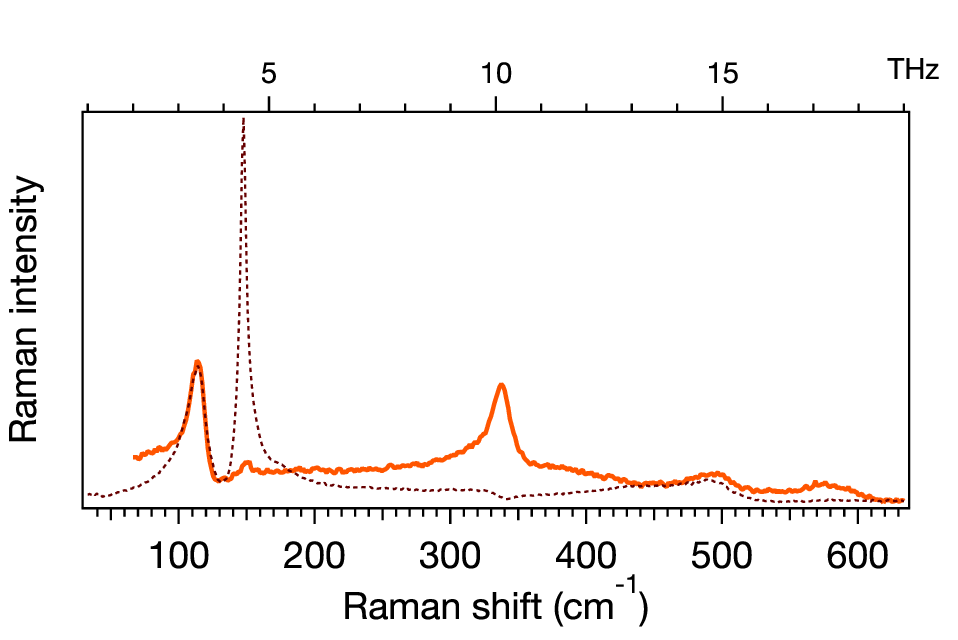}
\caption{\label{Raman} 
Raman scattering spectrum (solid curve) of YBa$_2$Cu$_3$O$_{7-\delta}$ measured at 785 nm in the back-scattering geometry. The FFT spectrum obtained from the NIR transient reflectivity measurement is shown by the dotted curve for comparison. The two spectra are normalized by the height of the Ba peak at 115 cm$^{-1}$.}
\end{figure}

Solid curve in Fig.~\ref{Raman} shows the spontaneous Raman scattering spectrum of YBa$_2$Cu$_3$O$_{7-\delta}$ measured at 785 nm.  The spectrum is consistent with those reported in previous studies on single crystals measured in the $xx$ or $yy$ polarization configurations at room temperature  \cite{Kulakovskii1988, Burns1988, Liu1988, Thomsen1988, Slakey1989, Friedl1991}.  
Comparison of the Raman spectrum with the FFT spectrum of the time-domain oscillation, shown by a dotted curve in Fig.~\ref{Raman}, reveals a striking difference.  Whereas the Ba mode at 3.5 THz ($\sim$115 cm$^{-1}$) is pronounced in both Raman and FFT spectra, the Cu(2) mode at 4.5 THz ($\sim$150 cm$^{-1}$), which had the largest FFT peak height, is very weak in the Raman spectrum. By contrast, the out-of-phase O(2,3) bending mode at 10 THz ($\sim$340 cm$^{-1}$), which was small in the FFT spectrum, is almost as intense as the Ba mode in the Raman spectrum.  

The stark contrast in the relative phonon amplitudes/intensities suggests an enhancement of the coherent Cu(2) phonon via a non-Raman generation mechanism, which is  elaborated for the general case in Appendix~\ref{DECP}.
For the Cu(2) mode of YBa$_2$Cu$_3$O$_{7-\delta}$ we propose a DECP mechanism associated with the ``buckling" of the CuO$_2$ plane. 
The asymmetric environment (Ba$^{2+}$ above, Y$^{3+}$ below) of the CuO$_2$ plane induces a built-in electric field perpendicular to the plane \cite{Devereaux1995, Opel1999}.  This leads to a slight separation of the plane formed by the Cu(2) ions and that by the O(2,3) ions, or buckling, instead of a flat CuO$_2$ plane.  The ionic motions associated with the buckling are the Cu(2) stretch and in-phase O(2,3) bend along the c direction, which can couple with static 
charge transfer 
between Cu(2) and O(2,3).  In the current normal-state study we expect both NIR and NUV pump pulses to induce an ultrafast charge transfer within the buckled CuO$_2$ plane between Cu(2) and O(2,3).  This would modify the equilibrium position of the Cu(2) ions in the c direction and thereby give a DECP-like driving force to the coherent Cu(2) mode at 4.5 THz.

The generation mechanisms of coherent phonons can be also inferred, to a certain degree, based on their initial phases $\phi_j$ \cite{Riffe2007, Watanabe2019, Merlin1997, Zeiger1992}.   
If the electronic excited state is short-lived or virtual, e.g. in non-resonant ISRS, the driving force can be approximated by a $\delta$-function of time (impulsive limit), and the ionic oscillation should follow a sine function time with $\phi_j=0$, as described in Appendix~\ref{ISRS}. If the excited state lives sufficiently long, e.g. in DECP, and the driving force can be approximated by a Heaviside step-function (displacive limit), the ionic oscillation would follow a cosine function, i.e., $\phi_j=\pi/2$, as described in Appendix~\ref{DECP}.
We find the Cu(2) oscillation component (red curves in Fig.~\ref{fit}a,b) to be nearly a cosine function of time ($\phi_\text{Cu}\simeq\pi/2$). 
The observation is consistent with our interpretation that the Cu(2) motion is driven efficiently via DECP mechanism.
In contrast, the phase of the Ba oscillation (black dotted curves) is closer to zero (sine-like), suggesting its coupling to an excited state with a shorter lifetime. 
This 
supports that the coherent Ba mode is generated predominantly via ISRS as a result of its weaker coupling with the in-plane charge transfer.  We note that in the present study we cannot determine the initial phases of the high-frequency modes because the uncertainty in the position of the zero time-delay (shaded gray in Fig.~\ref{fit}a,b) is too large for determining their initial phases.  

The low spectral weight of the 10-THz mode in the FFT spectrum represents another deviation from the ISRS generation mechanism; the coherent out-of-phase O(2,3) phonon is suppressed despite its large polarizability components ($a_{xx}, a_{yy}$).  Previous theoretical and Raman studies \cite{Devereaux1995, Opel1999} argued that the out-of-phase O(2,3) mode has a large Raman cross section because of its strong coupling to the static charge transfer between the O(2) and O(3) in the buckled CuO$_2$ plane.
We speculate that the photo-induced in-plane charge transfer between O(2,3) and Cu(2), which is responsible for the displacive excitation of the Cu(2) mode, could weaken the static charge transfer between the O(2) and O(3), and thereby reduces 
coherent out-of-phase O(2,3) motion.
We note that in the present pump-probe experiments we detect the spatial average of the contributions from orthogonal twin domains, since our laser spot is considerably larger than the typical twin domain size.  This does not affect the transient reflectivity signals of the 10-THz mode as well as other $A_g$ modes in the present polarization configuration, as long as they are generated via ISRS, as discussed in Appendix~\ref{detect}. On the other hand, \emph{if} the 10-THz mode is predominantly generated via non-Raman  (displacive) mechanism that is independent of the polarization of the incident pump light, the reflectivity modulation averaged over multiple twin domains could vanish.

We also note that in our Raman spectrum in Fig.~\ref{Raman} there is an additional weak peak at 577 cm$^{-1}$.  Previous Raman studies also reported a broad peak at a similar frequency \cite{McCarty1990, Krol1987, Liu1988, Altendorf1993, Hong2010}, though calculations predicted no Raman-allowed fundamental phonon mode in this frequency range \cite{Liu1988}.  It was attributed either to the $B_{2g}$ phonon mode arising due to the polarization leakage \cite{McCarty1990} or to a defect-activated O(1) stretching along the b axis ($B_{2u}$) based on its dependence on light polarization and oxygen deficiency \cite{Krol1987, Liu1988, Altendorf1993, Hong2010}.  
In our Raman spectra the intensity of the 577 cm$^{-1}$ peak varies considerably from spot to spot on the sample, which is consistent with the latter assignment associated with local defects.  \footnote{This mode is not distinct in our time-resolved measurements in the back-reflection geometry, but becomes more visible in the oblique-incidence geometry depending on the spot and on the pump/probe polarizations (see the black line in Fig.~\ref{obl}b).} 

\subsection{Oblique-incidence NIR pulse excitation}

\begin{figure}
\includegraphics[width=0.475\textwidth]{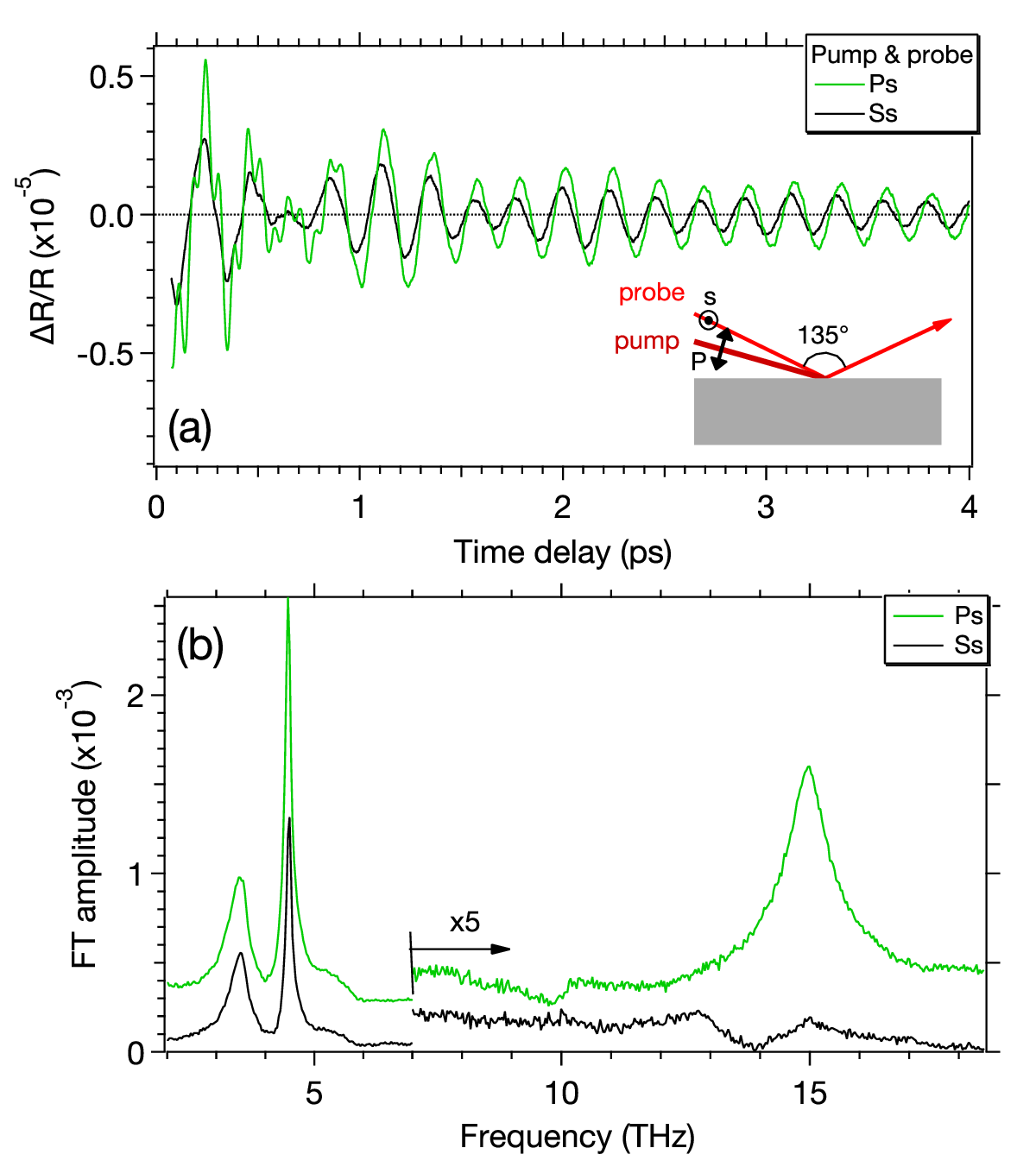}
\includegraphics[width=0.4\textwidth]{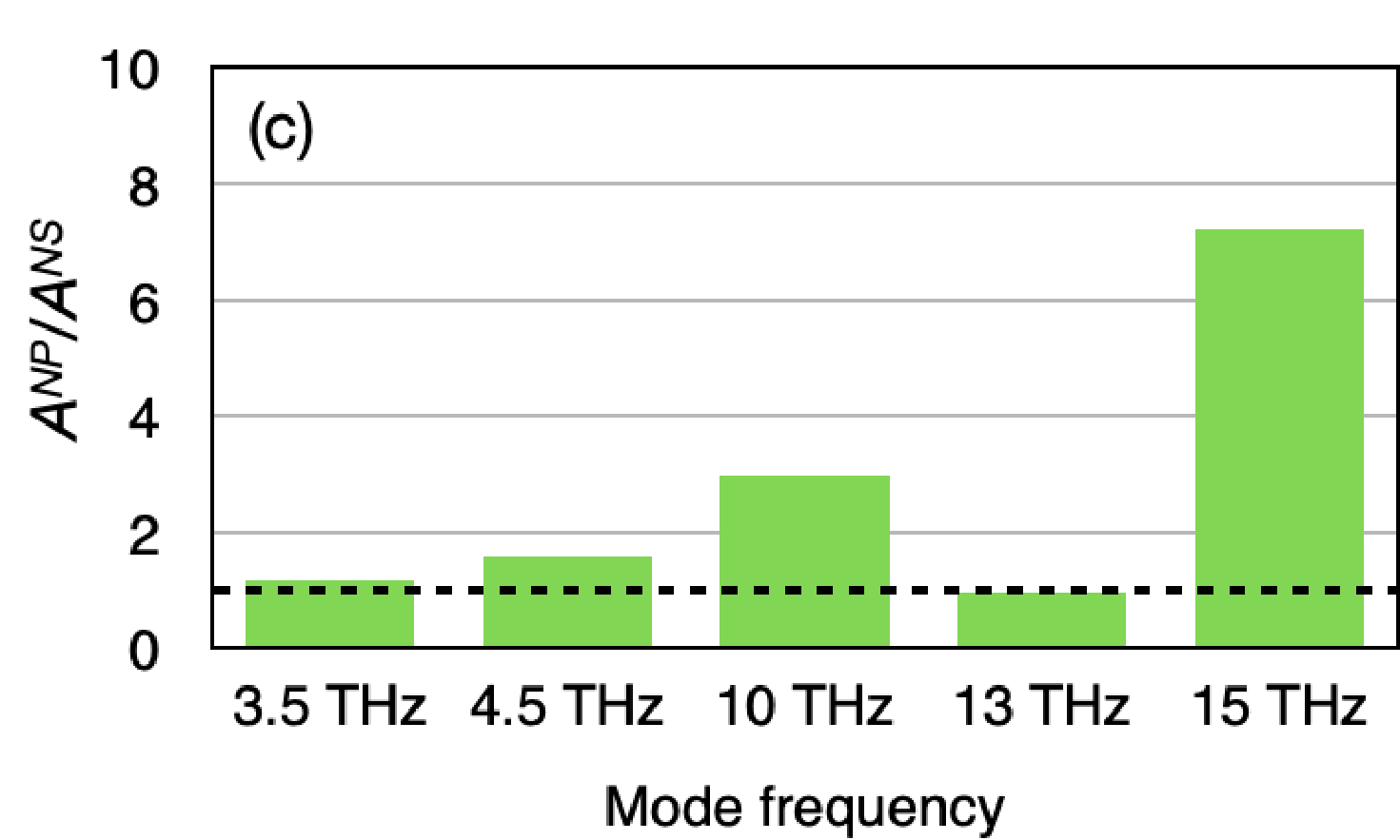}
\caption{\label{obl} (a,b) Oscillatory part of the transient reflectivity in YBa$_2$Cu$_3$O$_{7-\delta}$ (a) and its FFT spectrum (b) obtained in oblique-incidence NIR configuration with differently polarized pump and s-polarized probe.  Incident pump fluence is 5.1 $\mu$J/cm$^2$.  Photoexcited carrier density is $1.6\times10^{18}$ and $1.9\times10^{18}$ cm$^{-3}$ for s- and p-polarized pump.  Inset to (a) schematically illustrates the experimental geometry with p-polarized pump and s-polarized probe as an example.  The traces in (b) are offset for clarity. (c) Normalized coherent phonon amplitude ratio between p- and s-polarized pump, given by Eq.~(\ref{NR}).  Dashed line indicates $A^{NP}/A^{NS}=1$.
}
\end{figure}

So far we have discussed the transient reflectivity of YBa$_2$Cu$_3$O$_{7-\delta}$ in near normal-incidence geometry, in which the pump and probe beams are polarized in the ab plane. We further investigate coherent phonons with the NIR pulses in the 135$^\circ$-reflection geometry, as schematically illustrated in the inset of Fig.~\ref{obl}a. In this configuration s-polarized light, whose polarization component is only within the ab plane, induces the $A_1$ transition and gives the ISRS driving force described by Eq.~(\ref{force3}). By contrast,  p-polarized incident light has a polarization component along the c axis and therefore induce the $B_1$ transition, as shown in Fig.~\ref{Band}c,d.  Thus hhe ISRS generation would  be dominated by the $zz$ component of the polarizability:
\begin{eqnarray}\label{force4}
F^\textrm{ISRS}_z(t)&=&\big[(a_{xx}\cos^2\theta +a_{yy}\sin^2\theta)\cos^2\varphi\nonumber\\
&+&a_{zz}\sin^2\varphi\big]|E|^2\nonumber\\
&=&[0.15(a_{xx}\cos^2\theta+a_{yy}\sin^2\theta)+0.85a_{zz}]|E|^2,\nonumber\\
\end{eqnarray}
with $\varphi=135^\circ/2$ being the angle of incidence.

Figure~\ref{obl}a,b compares the oscillatory part of the transient reflectivity and its FFT spectrum obtained with differently polarized pump and s-polarized probe lights. 
As expected, the signal measured with s-polarized pump and probe (Ss), shown with black curves, is very similar to that obtained in the normal-incidence geometry (Fig.~\ref{TDFT800_400}).
Switching the pump to p-polarization while maintaining the probe polarization (Ps) leads to a drastic enhancement in the amplitude of the 15-THz O(4) mode and a moderate enhancement in other $A_g$ phonon modes, as shown with green curves.  
We obtain the amplitudes $A_j$, the frequencies $\nu_j$ and the dephasing rates $\Gamma_j$ of the coherent phonons with the same fitting procedure as described in Sect.~\ref{NormalIncidence}, whose results are shown 
in Appendix~\ref{AE}.  The frequencies obtained from the p-polarized pump are listed in the third column of Table~\ref{CPAg}.

The overall enhancement in the modes' amplitudes with p-polarized pump can be explained in terms of the lower reflectivity and the larger absorbed energy density for the p-polarized light than the s-polarized light. 
Indeed, if we normalize the phonon amplitudes $A_j$ by the absorbed light density $\alpha(1-R)$ for the respective polarization, we obtain the ratio between the normalized amplitudes:
\begin{equation}\label{NR}
\dfrac{A_j^{NP}}{A_j^{NS}}\equiv\dfrac{A_j^P/[\alpha_p (1-R_p)]}{A_j^S/[\alpha_s (1-R_s)]},
\end{equation}
that is reasonably close to unity, as shown in Fig.~\ref{obl}c, with an obvious exception of the O(4) mode at 15 THz. 

The pronounced enhancement of the 15-THz mode indicates significantly larger driving force for pump light polarized along the c axis than along the ab axes.  This observation can be explained qualitatively within the framework of the ISRS generation, since previous Raman studies \cite{Krol1987, Liu1988, Kulakovskii1988, Altendorf1991} reported larger polarizability along the c axis  ($a_{zz}\gg a_{xx}=a_{yy}$). Visible light polarized along the c axis induces the transition from the Cu(1)-O(1)-O(4) band to the unoccupied Cu(1)-O(1) chain band, labeled by $B_1$ in Fig.~\ref{Band}d. The zz polarizability component of the 500-cm$^{-1}$ (15-THz) mode is large because the O(4) displacement along the c axis modulates the O(4)-Cu(1) and O(4)-O(1) distances and hence of the O(4)-O(1) hopping.  
The earlier Raman studies  \cite{Krol1987, Liu1988, Kulakovskii1988, Altendorf1991} reported a similar, though weaker, enhancement for the 13-THz mode, which was explained in terms of the mode mixing with the O(4) vibration \cite{Heyen1990}.  In the present study we see no such enhancement, however.  It is therefore possible that the generation of the 15-THz mode with p-polarized light is additionally enhanced by its stronger coupling with the optical transition induced by the p-polarized NIR light via a DECP-like mechanism.

The larger amplitudes in this configuration enables us to discuss the parameters of the high-frequency modes more confidently.  Specifically, the frequency and the dephasing rate of the 15-THz mode are obtained to be $\nu_{15}=14.9$~THz (497~cm$^{-1}$) and $\Gamma_{15}=2.4$~ps$^{-1}$.  Our frequency is comparable with that reported in the earlier Raman studies (494-504 cm$^{-1}$ at 300 K) \cite{McCarty1990, McCarty1991, Altendorf1991, Altendorf1993, Limonov1998}. Our dephasing rate is also consistent with the previously reported Raman linewidth (FWHM=25-28 cm$^{-1}$ at 300 K) \cite{McCarty1991, Altendorf1991, Altendorf1993, Limonov1998}, whose temperature-dependence revealed anharmonic decay into low-frequency phonons being the dominant decay channel.
In the present study we detect no noticeable photo-induced frequency shift in the O(4) mode, in contrast to the recent time-resolved Raman study that employed more than two-orders-of-magnitude higher excitation density \cite{Pellatz2021}.
The in-phase O(2,3) mode at 13 THz decays almost as fast ($\Gamma_{13}=2.1$~ps$^{-1}$), in agreement with the linewidth (FWHM$\gtrsim$30 cm$^{-1}$) reported in earlier studies \cite{Altendorf1991, Altendorf1993}.

Finally we note that the observed enhancement of the 15-THz mode for the pump polarized along the c axis is in an apparent contrast to the polarization-dependence study of undoped single-layer cuprate La$_2$CuO$_4$ \cite{Baldini2020}, where the La--apical O vibrations along the c axis were suppressed under pump$\parallel$c compared to pump$\parallel$a.  This is because at the photon energy of 3.1 eV the pump$\parallel$a leads to an efficient excitation across the charge-transfer gap of La$_2$CuO$_4$ that strongly couples with the lattice degree of freedom, whereas along the c axis the crystal is more insulating.  
In the present study on YBa$_2$Cu$_3$O$_{7-\delta}$ we have not examined the $B_2$ transition at $\sim$3 eV directly by performing the NUV oblique incidence measurement. 
We anticipate, however, that the $B_2$ excitation would involve an analogous charge transfer channel as the $B_1$ and would give rise to a similar result as in the case of $A_1$ and $A_2$ transitions. 

\section{CONCLUSIONS}

We have investigated the nature of electron-phonon coupling in optimally doped YBa$_2$Cu$_3$O$_{7-\delta}$ at room temperature under NIR and NUV  photo-excitations.  Our ultrashort laser pulses have allowed us to time-resolve the three high-frequency phonon modes involving O(2,3) and O(4), in addition to the previously reported low-frequency Ba and Cu(2) modes.  Comparison of the relative coherent phonon amplitudes with the relative Raman intensities has revealed the mode-dependent coupling with the electronic excitation.  The coherent Cu(2) mode is enhanced significantly via the displacive driving mechanism facilitated by the ultrafast in-plane charge transfer between the Cu(2) and O(2,3) ions.  The coherent out-of-phase O(2,3) mode, by contrast, is suppressed drastically.
The excitation of other $A_g$ modes can qualitatively be explained within the framework of the impulsive Raman generation. %
Our results present a good example for discussing coherent phonon generation in a complex crystal, in which the electronic excitation can be relatively localized within the unit cell and can couple with different lattice modes in qualitatively different manners.  Similar time-resolved experiments with ultrashort pulses would extend access to high-frequency phonon modes of cuprates and correlated materials in general. Tracking the temperature- and doping-dependences of the coherent phonon response could further reveal which of the phonon modes display particularly strong anomalies and the related contribution to superconductivity.  

\begin{acknowledgments}
We would like to thank Ece Uykur for fruitful discussions. This work was supported by the Deutsche Forschungsgemeinschaft (DFG, German Research Foundation) TRR 288-422213477 (“Elasto-Q-Mat”, Project B08) and the Swiss National Science Foundation (SNSF) through project 200020-172611.
\end{acknowledgments}

\appendix

\section{Theory for generation and detection of coherent phonons}\label{theory}

\subsection{Raman generation of coherent phonons}\label{ISRS}

The electron-phonon coupling behind the generation of coherent phonons can depend on the material as well as the pump photon energy.  
When the photon energy is below the fundamental bandgap (e.g. in a semiconductor), the coherent phonons can be driven via impulsive stimulated Raman scattering (ISRS) mechanism \cite{Dhar1994}, in which a broadband femtosecond optical pulse offers multiple pairs of photons required for the stimulated Raman process.  
In this case the driving force $F$ depends on the polarization of the pump electric field $E$ through a third-rank Raman tensor $\mathfrak{R}_{jkl}\equiv(\partial\chi/\partial Q)_{jkl}\equiv a_{kl}^{(j)}$ \cite{Merlin1997, Dekorsy, Mann2015}:
\begin{equation}\label{A1}
F^\textrm{ISRS}_j(t)=\mathfrak{R}_{jkl}E_k(t)E_l(t),
\end{equation}
where $j, k, l$ are Cartesian coordinates. This equation implies that the pump electric field acts directly upon the ions in the crystal through the Raman tensor, whose components are described by the polarizability $a_{kl}^{(j)}$:
\begin{equation}\label{A2}
\mathfrak{R}(j)= \begin{pmatrix}
a_{xx} & a_{xy} & a_{xz}\\
a_{yx} & a_{yy} & a_{yz} \\
a_{zx} & a_{zy} & a_{zz}
\end{pmatrix}.
\end{equation}

For the light field $E(t)$ whose duration is shorter than the phonon period $\Omega^{-1}$, this driving force can be approximated with a $\delta$-function of time and would induce a ``coherent" displacement  {\bf Q} in the collective ionic position in a crystal \cite{Dekorsy}:
\begin{equation}\label{A3}
\mu\Big(\dfrac{\partial^2 {\bf Q}}{\partial t^2}+2\Gamma \dfrac{\partial {\bf Q}}{\partial t}+\Omega^2 {\bf Q}\Big)={\bf F}^\text{ISRS}(t),
\end{equation}
where $\Omega\equiv2\pi\nu$ is the angular frequency. $\mu$ and $\Gamma$ denote the reduced lattice mass and a phenomenological damping constant.  The solution to Eq.~(\ref{A3}) is given by a damped harmonic function:
\begin{equation}\label{A4}
{\bf Q}(t)={\bf Q}_0\exp(-\Gamma t)\sin(\Omega t+\phi).
\end{equation}
The amplitude ${\bf Q}_0$ is proportional to the driving force ${\bf F}^\text{ISRS}$, and hence the polarizability component $a_{kl}$ as described by eqs.~(\ref{A1}) and (\ref{A2}).  For the non-resonant ISRS we can assume  a boundary condition ${\bf Q}(t=0)=0$, and the resultant oscillation is expressed by a sine function of time with the initial phase $\phi=0$.

For the $A_g$-symmetry phonons of orthorhombic YBa$_2$Cu$_3$O$_{7-\delta}$, whose Raman tensor is given by Eq.~(\ref{Rtensor}), Eq.~(\ref{A1}) reduces to:
\begin{equation}\label{A5}
F^\textrm{ISRS}_z(t)=a_{xx}|E_x|^2+a_{yy}|E_y|^2+a_{zz}|E_z|^2.
\end{equation}
In the back-reflection geometry from ab plane, Eq.~(\ref{A5}) reduces further to Eq.~(\ref{force3}).  By representing the polarization angle of the pump relative to the a axis with $\theta$ we can express the driving force as:
\begin{equation}\label{A6}
F^\textrm{ISRS}_z(t)=(a_{xx}\cos^2\theta+a_{yy}\sin^2\theta)|E|^2.
\end{equation}

\begin{figure}
\includegraphics[width=0.4\textwidth]{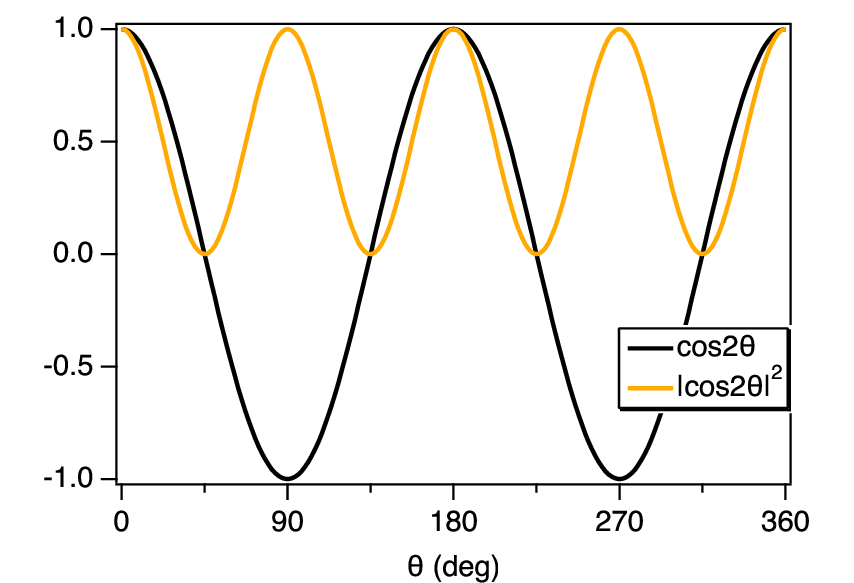}
\caption{\label{theta}  $\theta$-dependences of the functions included in eqs.~(\ref{A7}) and (\ref{A15}).}
\end{figure}

\begin{figure}
\includegraphics[width=0.35\textwidth]{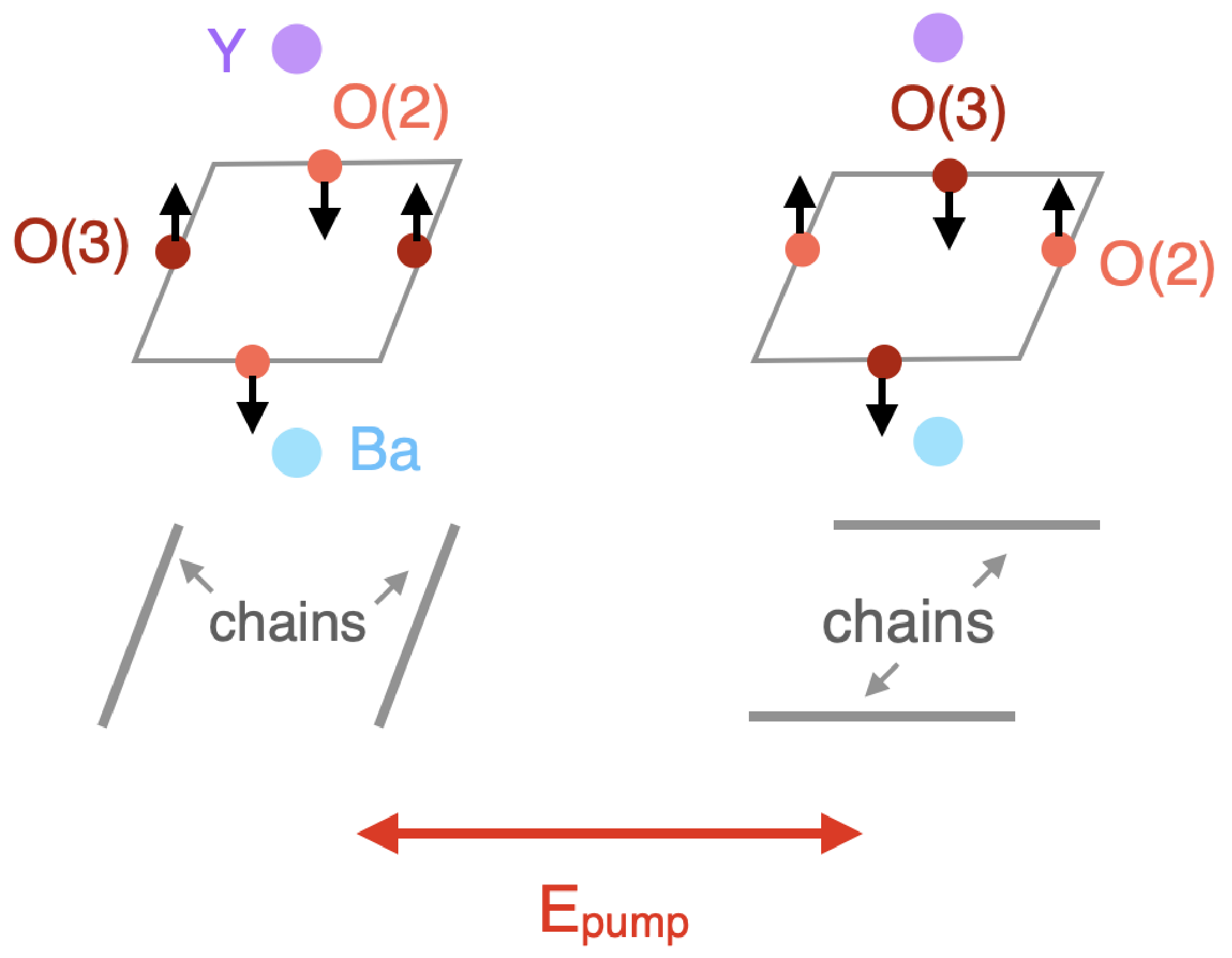}
\caption{\label{Raman2} 
Unit cells of the CuO$_2$ plane in orthogonal twin domains of YBa$_2$Cu$_3$O$_7$ showing the ionic displacements corresponding to the out-of-phase O(2,3) modes at 10 THz.  ISRS mechanism induces driving forces with the opposite directions between pump$\parallel$a (left) and pump$\parallel$b (right) domains.
}
\end{figure}

Among the five $A_g$ modes of YBa$_2$Cu$_3$O$_{7-\delta}$, only the out-of-phase O(2,3) mode has an asymmetric $B_{1g}$-like character ($a_{yy}=-a_{xx}$), whereas all the other $A_g$ modes are symmetric ($a_{yy}=a_{xx}$) \cite{Liu1988, Kulakovskii1988, McCarty1990}.  For the four symmetric modes Eq.~(\ref{A6}) reduces to 
\begin{equation}
F^\text{ISRS}=a_{xx}|E|^2
\end{equation}
regardless of $\theta$.  For the out-of-phase O(2,3) mode, however, the driving force depends on $\theta$ as:
\begin{equation}\label{A7}
F^\textrm{ISRS}_z(t)=a_{xx}|E|^2\cos2\theta,
\end{equation}
whose angular-dependence is shown with a black curve in Fig.~\ref{theta}.  
In the twinned YBa$_2$Cu$_3$O$_{7-\delta}$ crystal used in the present study, this would lead to ionic motions in the opposite directions between the two orthogonal domains (pump$\parallel$a and pump$\parallel$b), as illustrated in Fig.~\ref{Raman2}.

\subsection{Non-Raman generation of coherent phonons}\label{DECP}

When the photon energy exceeds the bandgap (or in a metallic system), the Raman process involved in the excitation and detection of coherent phonons can be resonantly enhanced in the similar manner as in spontaneous Raman scattering  \cite{Stevens2002, Cardona1982}.  In addition, non-Raman mechanisms  \cite{Zeiger1992, Dekorsy} can contribute to the coherent phonon generation. 
So-called displacive excitation of coherent phonons (DECP) was originally proposed for the ionic motion along the trigonal axis of rhombohedral Bi and Sb  \cite{Zeiger1992}, which is associated with the Peierls instability. Transient depletion field screening (TDFS) proposed for the longitudinal optical (LO) phonons in the depletion layer of doped compound semiconductors \cite{Dekorsy, Ishioka2015} can be considered as a variation of DECP.
The DECP driving force is given by the dependence of the equilibrium coordinate $Q_E(t)$ on excited carrier density $N$ (or electronic temperature $T_e$), which is assumed to be linear:
\begin{equation}\label{A8}
Q_E(t)\propto N(t).
\end{equation}
The equation of motion is taken as:
\begin{equation}\label{A9}
\mu\Big[\dfrac{\partial^2 Q(t)}{\partial t^2}+2\Gamma \dfrac{\partial Q(t)}{\partial t}+\Omega^2 (Q(t)-Q_E(t))\Big]=0.
\end{equation}
Equations~(\ref{A8}) and (\ref{A9}) imply that the DECP driving force is applied to the ions through the interaction with photoexcited carriers, in contrast to the electric field acting directly on the ions in ISRS.  The solution to Eq.~(\ref{A9}) is given by a damped harmonic function:
\begin{equation}\label{A10}
Q(t)=Q_E \exp(-\Gamma t)\sin(\Omega t+\phi).
\end{equation}
When the photoexcited carriers decay fast enough compared with the phonon period, the DECP driving force can be approximated by a $\delta$-function of time (impulsive limit), and the induced ionic displacement can be expressed by the similar solution as the ISRS (Eq.~(\ref{A4})) that follows a sine function of time, i.e., $\phi=0$ or $\pi$.  When the carriers are created fast and live sufficiently long, the driving force can be approximated by a Heaviside step-function (displacive limit), and the ionic oscillation would follow a cosine function, i.e., $\phi=\pi/2$.  For an intermediate carrier lifetime the initial phase of the displacement oscillation would take a value between the two extreme cases \cite{Zeiger1992}.

\subsection{Detection of coherent phonons by transient reflectivity}\label{detect}

Once generated by an ultrashort pump pulse, whether via ISRS or DECP, the coherent ionic displacement $Q$ induces a change in the complex refractive index, $\Delta n$, and thereby in the reflectivity, $\Delta R$.  In a first-order approximation the reflectivity change can be rewritten using the susceptibility $\chi$ \cite{Merlin1997, Dekorsy, Mann2015}:
\begin{equation}\label{A11}
\Delta R=\frac{\partial R}{\partial n}\Delta n\simeq \sum_{pq}\frac{\partial R}{\partial \chi}\Bigl(\frac{\partial \chi}{\partial Q}\Bigr)_{jpq} Q_j {e}_{p} {e}_{q}.
\end{equation}
Here $(\partial \chi/\partial Q)_{jpq}\equiv  \mathfrak{R}_{jpq}\equiv a_{pq}^{(j)}$ is the first-order Raman tensor given by Eq.~(\ref{A2}). ${e}_{p}$ and ${e}_{q}$ denote the $p$ and $q$ components of the unit vector $\bf{E}/|\bf{E}|$ of the \emph{probe} light field, with $p$ and $q$ denoting the Cartesian coorinates.  
Equation~(\ref{A11}) implies that the displacement $Q$ induces the reflectivity change $\Delta R$ through the Raman polarizability components $a_{pq}$ selected by the probe light polarization. 

For coherent phonons generated via ISRS, eqs.~(\ref{A1}) and (\ref{A11}) would imply that the phonon-induced $\Delta R$ is proportional to the product of $a_{kl}$ defined by the pump polarization and $a_{pq}$ defined by the probe polarization.  In the case of the $A_g$ modes pumped and probed within the ab plane the polarization angle-dependence can be expressed by:
\begin{equation}\label{A12}
\Delta R\propto (a_{xx}\cos^2\theta+a_{yy}\sin^2\theta)(a_{xx}\cos^2\theta'+a_{yy}\sin^2\theta'),
\end{equation}
with $\theta$ and $\theta'$ denoting the polarization angles of the pump and probe.  For the four symmetric ($a_{yy}=a_{xx}$) $A_g$ modes of YBa$_2$Cu$_3$O$_{7-\delta}$, Eq.~(\ref{A12}) would reduce to:
\begin{eqnarray}\label{A13}
\Delta R&\propto&a_{xx}^2 (\cos^2\theta+\sin^2\theta)(\cos^2\theta'+\sin^2\theta')\nonumber\\
&=&a_{xx}^2.
\end{eqnarray}
%
%
%
For the asymmetric ($a_{yy}=-a_{xx}$) out-of-phase O(2,3) mode Eq.~(\ref{A12}) would lead to:
\begin{eqnarray}\label{A15}
\Delta R&\propto&a_{xx}^2 (\cos^2\theta-\sin^2\theta)(\cos^2\theta'-\sin^2\theta')\nonumber\\
&=&\left\{\begin{array}{ll}
a_{xx}^2 \cos^22\theta&\text{for}\; \theta'-\theta=0\\
-a_{xx}^2 \cos^22\theta&\text{for}\; \theta'-\theta=\pi/2.
\end{array}
\right.
\end{eqnarray}
The angular-dependence for the parallel polarizations ($ \theta'-\theta=0$) is shown with a yellow curve in Fig.~\ref{theta}.  
Equation~
(\ref{A15}) indicates that, for a fixed relative angle between the pump and probe (e.g. $\theta'-\theta=0$), the reflectivity oscillations from orthogonal twin domains ($\theta=0$ and $\pi/2$) would add up if the detection is averaged over multiple domains.

\subsection{Spontaneous Raman scattering intensity}\label{SpRaman}

(Spontaneous) Raman scattering can be regarded as a radiation from an optically-induced polarization in a crystal.  The polarization {\bf P} induced by an electric field {\bf E} can be written in the form ${\bf P}=\chi {\bf E}$ with the electric susceptibility $\chi$ \cite{YuCardona}.  
The susceptibility, in turn, can be written as a function of ionic displacement ${\bf Q}={\bf Q}_0\sin(\Omega t+\phi)$. In the first-order approximation the intensity of the scattered radiation will depend on the polarizations of the incident and scattered lights, ${\bf e}_i$ and ${\bf e}_s$, as:
\begin{equation}\label{A19}
I_s\propto|{\bf e}_i \cdot (\partial \chi/\partial {\bf Q})_0{\bf Q}_0 \cdot {\bf e}_s|^2.
\end{equation}
This formula implies that the scattered intensity is proportional to the polarizability component $a\equiv(\partial \chi/\partial {\bf Q}_0)$ \emph{squared}.  By introducing a unit vector ${\bf e}_Q ={\bf Q/|Q|}$ and the second rank Raman tensor $\mathfrak{R}=(\partial \chi/\partial {\bf Q}_0) {\bf e}_Q$ we can rewrite Eq.~(\ref{A19}) as:
\begin{equation}\label{A20}
I_s\propto|{\bf e}_i\cdot \mathfrak{R}\cdot{\bf e}_s|^2.
\end{equation}

Previous Raman studies on orthorhombic YBa$_2$Cu$_3$O$_{7-\delta}$ derived the Raman tensors of the phonon modes based on their polarization-dependences  \cite{Liu1988, Kulakovskii1988, McCarty1990, Krol1987, Liu1988, Kulakovskii1988, Altendorf1991}.  In the back-scattering geometry from the ab plane, five peaks were observed at $\sim$115, 150, 340, 440 and 500 cm$^{-1}$ in the $xx$ (${\bf e}_i, {\bf e}_s\parallel$a) and $yy$ (${\bf e}_i, {\bf e}_s\parallel$b) configurations  but vanished in the $xy$  (${\bf e}_i\parallel$a, ${\bf e}_s\parallel$b)  \cite{Liu1988, Kulakovskii1988, McCarty1990}.  This indicates the $A_g$ symmetry of the phonons, with their Raman tensor expressed in the form of  Eq.~(\ref{Rtensor}).  
In the parallel polarization the intensities of the four peaks at 115, 150, 440 and 500 cm$^{-1}$ were nearly independent of the polarization angle of the incident light with respect to the crystallographic axes, indicating $a_{xx}\simeq a_{yy}$.  Only the 340~cm$^{-1}$ peak exhibited qualitatively different dependence; it was maximum when the incident light is polarized along the a or b axis but vanished at 45$^\circ$.  In the cross polarization with the incident light polarized at $\theta=45^\circ$, only this mode was detected distinctly while the other four modes vanished.  From these observations the 340~cm$^{-1}$ mode was determined to be asymmetric ($a_{xx}=-a_{yy}$).

Further polarized Raman studies in the $zz$ configuration (${\bf e}_i, {\bf e}_s\parallel$c) determined the magnitude of $a_{zz}$ relative to $a_{xx}$ and $a_{yy}$ \cite{Krol1987, Liu1988, Kulakovskii1988, Altendorf1991}. In particular, the Raman intensities of the 500 and 440 cm$^{-1}$ modes were found to be much larger in the $zz$ configuration than in the $xx$ and $yy$, which implies the largest polarizability along the c axis ($a_{zz}\gg a_{xx}=a_{yy}$).  By contrast, the 340 cm$^{-1}$ mode vanished in the $zz$ configuration, indicating a vanishing polarizability ($a_{zz}\simeq 0$).

\section{Effect of continuous laser heating}\label{Heating}

\begin{figure}
\includegraphics[width=0.475\textwidth]{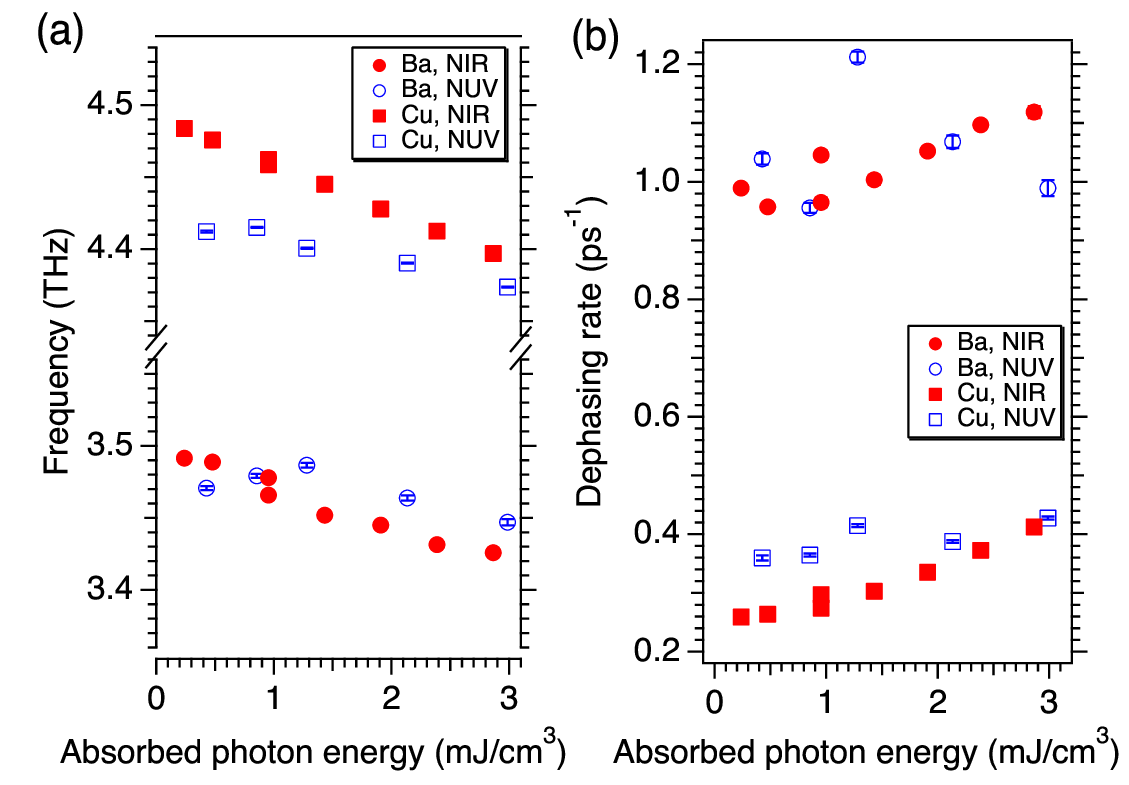}
\caption{\label{Power2}  Pump density-dependence of frequencies (a) and dephasing rates (b) of the coherent Ba and Cu(2) phonons obtained from fitting the NIR and NUV time-domain data to Eq.~(\ref{ddh}). }
\end{figure}

In the present study we used a Ti:sapphire oscillator with a high (80 MHz) repetition rate as the source for the pump-probe measurements. In this setup the effect of continuous laser heating at the laser-irradiated YBa$_2$Cu$_3$O$_{7-\delta}$ surface may not be neglected. 
A simple steady-state heat diffusion model \cite{heating, mihailovic1999b} predicts that the temperature increase is roughly proportional to fluence.  Using Eq.~(3) of Ref.\cite{mihailovic1999b} and the literature values for the anisotropic thermal conductivity \cite{Hagen1989} and for the optical constants \cite{Kircher1991} we estimate the steady state heating to be $\approx$ 170 K at the highest excitation fluence used in this experiment.

Continuous heating of this magnitude would affect the frequencies $\nu_j$ and the dephasing rates $\Gamma_j$ of the coherent phonons significantly.  Previous temperature-dependent Raman studies \cite{Altendorf1993, Macfarlane1987} reported frequency redshift and linewidth broadening for the out-of-phase O(2,3) and O(4) modes with increasing temperature above $T_c$.  These trends can be explained in terms of the bond softening and the enhanced phonon-phonon scattering at an elevated temperature. 
In the present study,  
the Ba and Cu(2) modes display similar frequency redshifts and enhancements in the dephasing rates with increasing fluence, as shown in Fig.~\ref{Power2}, supporting the laser-induced temperature rise in our experimental setup.

\section{Comparison of fitting to double and triple damped harmonic functions}\label{2vs3}

\begin{figure}[ht]
\includegraphics[width=0.475\textwidth]{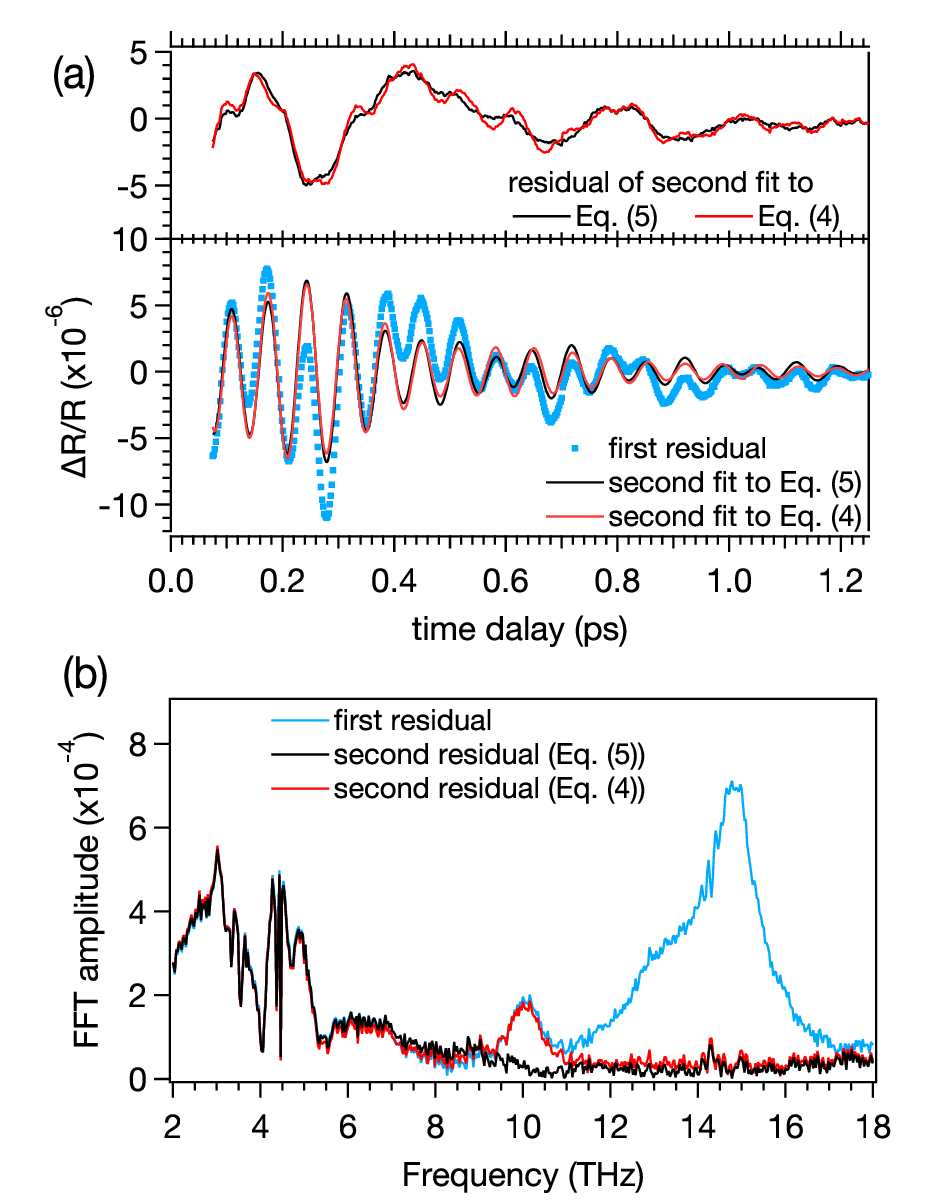}
\caption{\label{fit2} (a) Residual of the first fit of the NIR transient reflectivity to eq~(\ref{ddh}) (blue dots in lower panel) and its second fits to Eq.~(\ref{mdh}) using three-components at 10, 13, and 15 THz (black curve in lower panel) and using two components at 13 and 15 THz (red curve in lower panel).  The corresponding second residuals are also shown in the upper panel. (b) FFT spectra of the first residual (blue), the second residual using the tree components (red) and that using the two components (black).}
\end{figure}

Fitting of the NIR oscillatory transient reflectivity to Eq.~(\ref{ddh}) leaves a first residual shown with dots in the upper panel of Fig.~\ref{fit}a and in the lower panel of Fig.~\ref{fit2}a.  The first residual exhibits a beating pattern at an apparently higher frequency than the Ba and Cu(2) modes.  The FFT spectrum of the first residual, shown in Fig.~\ref{fit2}b, suggests the largest contribution from the 15-THz mode, but also those from the 13- and 10-THz modes.  Figure~\ref{fit3}a compares the fitting of the first residual to a triple damped harmonic function including the 10-, 13-, and 15-THz modes with that to a double damped harmonic function including 13- and 15-THz modes. 
The three-component function reproduces the first remnant almost perfectly in the frequency range between 8 and 18 THz, whereas the two-component function fails to fit the small contribution at 10 THz, as shown in Fig.~\ref{fit2}b.  In both approaches, the residual after the second fitting features a broad spectral structure extending from 2 to 5 THz, whose amplitude is 20 times weaker than the original signal.  The low-frequency residual indicates a small deviation of the Ba and Cu modes from damped harmonic oscillations in the first picosecond.

\section{Implication of the choice of the time-window on the FFT}\label{TimeWindow}

\begin{figure}
\includegraphics[width=0.475\textwidth]{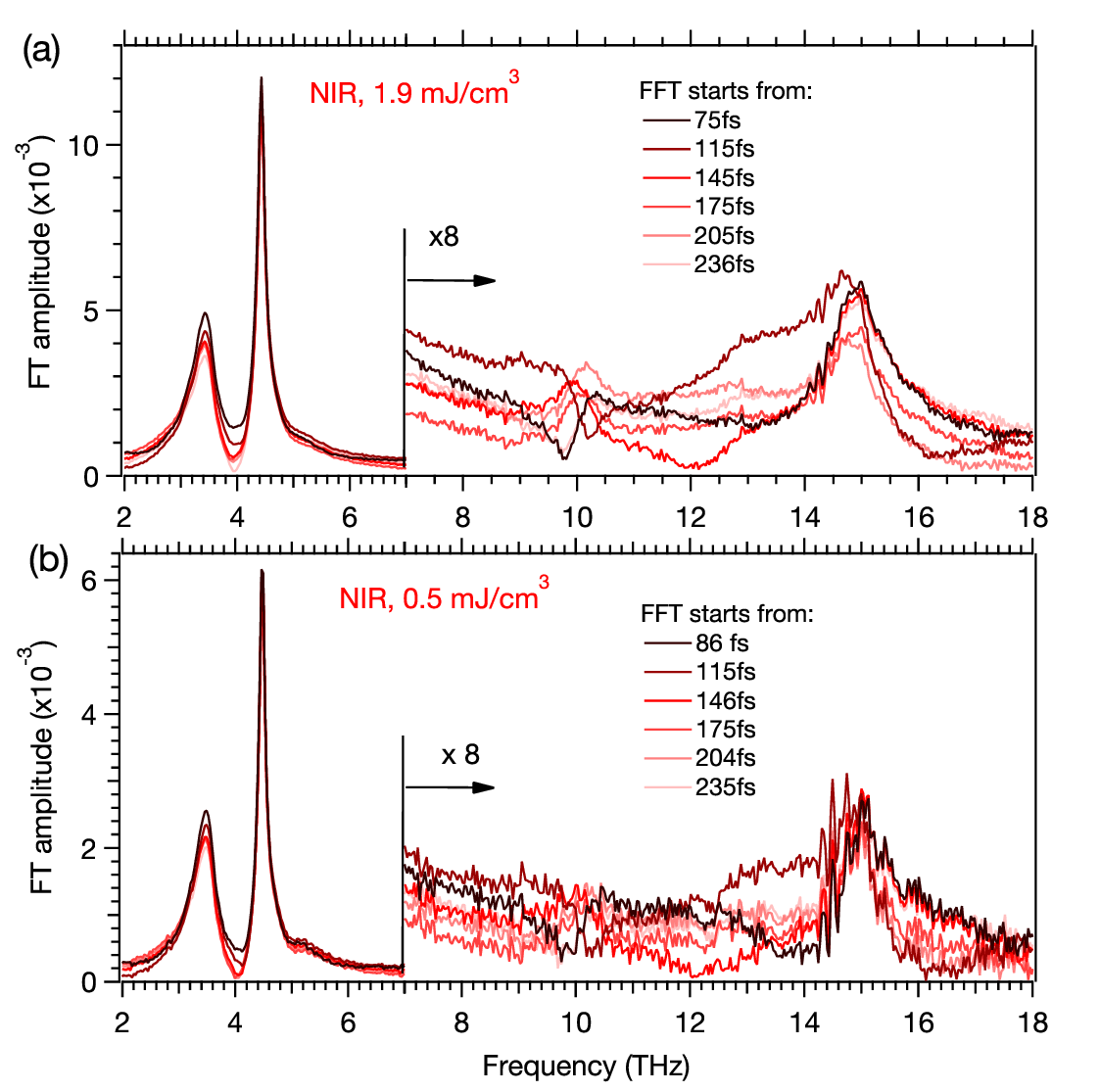}
\includegraphics[width=0.475\textwidth]{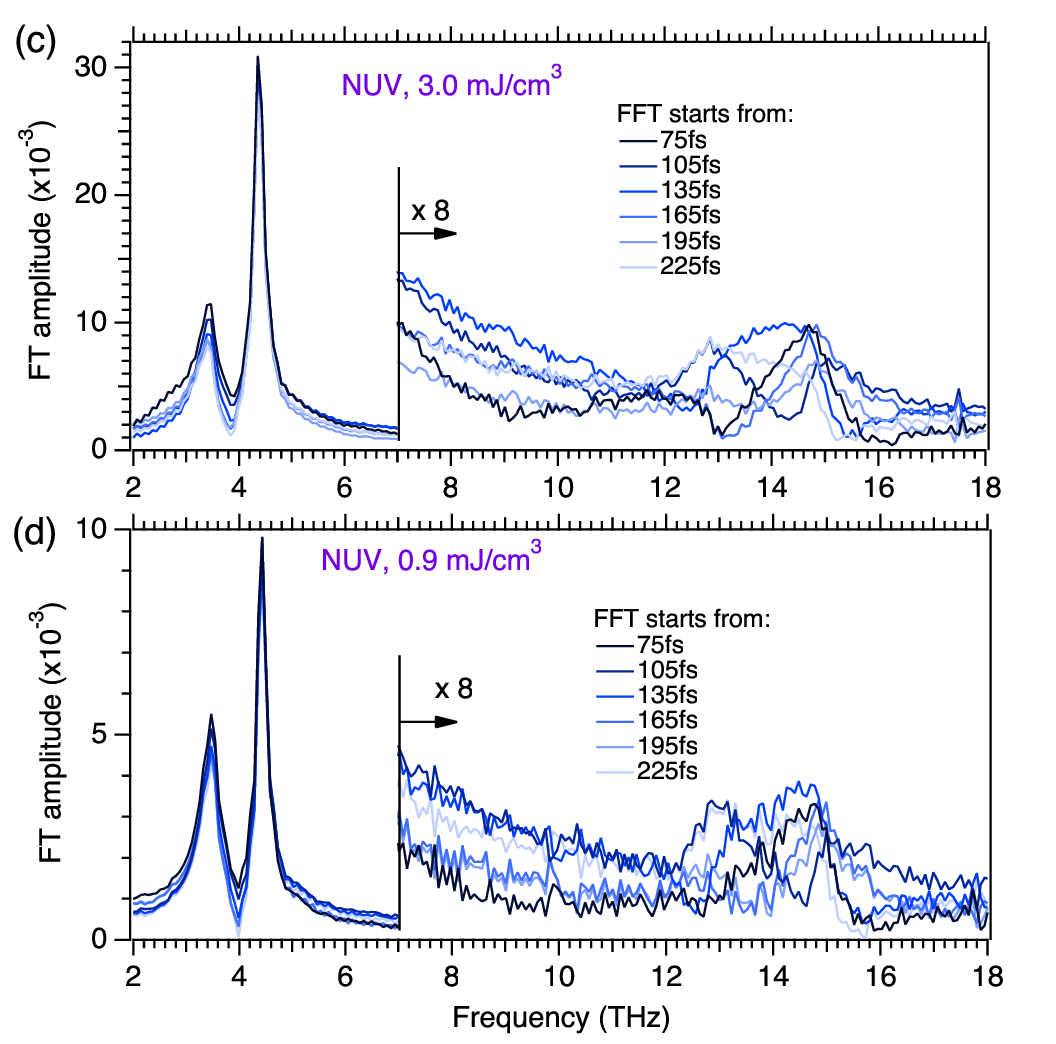}
\caption{\label{FFT_start} FFT spectra of the oscillatory transient reflectivity obtained from the NIR (a,b) and NUV (c,d) experiments at high (a,c) and low (b,d) pump fluences. The FFT are performed over time-windows starting from different initial time delays.}
\end{figure}

In the high-frequency region (8-18 THz) of the FFT spectra (Fig.~\ref{TDFT800_400}) the phonon modes appear as a complex structure of peaks and dips (negative peaks) overlapping each other, rather than isolated peaks.  The spectral lineshape in this frequency region actually depends on the temporal window employed for the FFT, as demonstrated in Fig.~\ref{FFT_start}.  For the normal-incidence NIR experiments (Fig.~\ref{FFT_start}a,b) we see the 15-THz mode is most intense and always appears as a peak, whereas the weaker 10- and 13-THz modes change their shape drastically and periodically from a peak to a bipolar (peak/dip) structure to a dip with shifting the starting time for the FFT.

Similar time window-dependence of the FFT spectral lineshape was previously reported for the weak $E_g$ mode of Bi appearing on the spectral tail of the intense $A_g$ mode \cite{Misochko2006}.
The periodic spectral evolution of the weaker phonon modes can be explained in terms of the interference among the phonon modes.  Because the oscillation amplitudes decrease with time, the lineshapes are determined essentially by the phase relation among the phonon modes (whether they are in-phase or out-of-phase) at the start of the time window.  It is also possible that the high-frequency tail of the intense Cu(2) mode is 
contributing to the interference, especially for the 10-THz mode.  This conjecture is supported by the flatter baseline and the more isolated lineshape of the 10-THz mode in the FFT spectra of the first residual, which is after subtraction of the two low-frequency modes in Fig.~\ref{fit2}b.  

For the NUV experiment we see essentially similar spectral evolution, except that the 10-THz mode is too week to be detected and that the 13- and 15-THz modes have comparable amplitudes.  As a consequence, not only the 13-THz mode but also the 15-THz mode changes its spectral shape drastically.

\section{Fitting to data obtained in oblique detection geometry}\label{AE}

\begin{figure}
\includegraphics[width=0.475\textwidth]{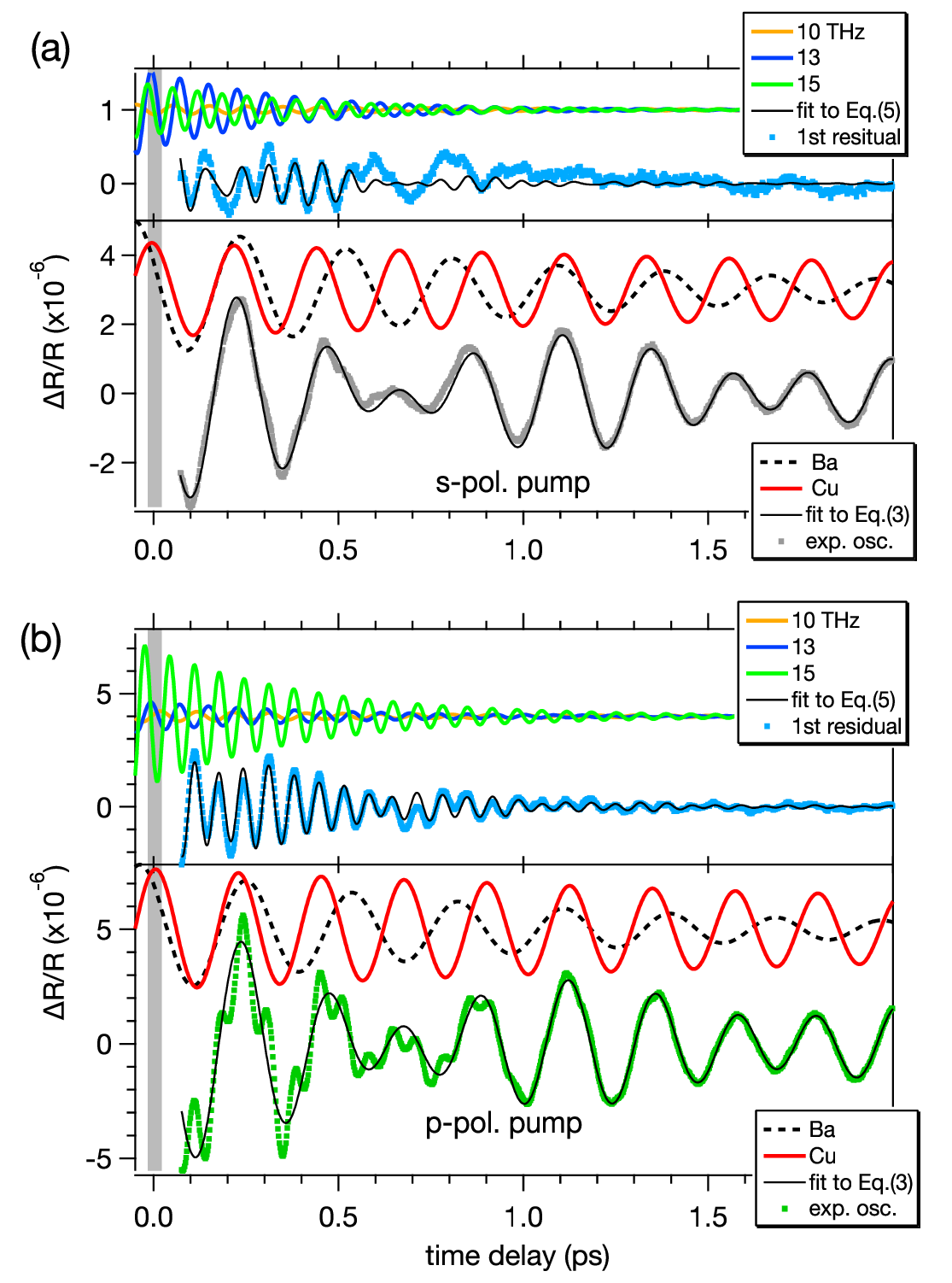}
\caption{\label{fit3} Oscillatory transient reflectivity (dots in bottom panels) obtained with s-polarized (a) and p-polarized (b) NIR pump 
in oblique-incidence geometry.  Black solid curve, dashed curve and red solid curve in bottom panels are fit to Eq.~(\ref{ddh}) and its Ba- and Cu-mode contributions, with the latter two being offset for clarity.  Dots in the upper panels represent the residual of the fitting shown in the bottom panels.  Black and colored curves are fit to Eq.~(\ref{mdh}) and contributions from the 10-, 13- and 15-THz modes, the latter three being offset for clarity. }
\end{figure}

The oscillatory transient reflectivity obtained from oblique-incidence NIR experiment, shown in Fig.~\ref{obl}a, are analyzed with the similar procedure to that employed in the normal-incidence experiments.  The fit results are shown in Fig.~\ref{fit3} for the s- and p-polarized pump.

\bibliographystyle{apsrev4-2}
\bibliography{YBCO}

\end{document}